\documentclass[%draft,published,
notoc,nohyper]{JHEP3} % 10pt is ignored!

\newcommand{\reseteqnum}{\setcounter{equation}{0}}
\newcommand{\nn}{\nonumber}
\newcommand{\eqn}[1]{(\ref{#1})}
\newcommand{\ovl}[1]{\overline{#1}}
\newcommand{\wt}[1]{\widetilde{#1}}
\newcommand{\wh}[1]{\widehat{#1}}
\newcommand{\p}{\partial}
\newcommand{\dslash}{\p\kern-1.2ex /}
\newcommand{\Dslash}{D\kern-1.5ex /}
\newcommand{\bphi}{{\overline{\phi}}}
\newcommand{\bPsi}{{\overline{\Psi}}}
\newcommand{\bpsi}{{\overline{\psi}}}

\newcommand{\brho}{{\overline{\rho}}}
\newcommand{\blambda}{{\overline{\lambda}}}
\newcommand{\hP}{{\widehat{\Pi}}}
\newcommand{\oP}{{\overline{\Pi}}}
\newcommand{\tPi}{{\widetilde{\Pi}}}
\newcommand{\otPi}{{\overline{\widetilde{\Pi}}}}
\newcommand{\tP}{{\widetilde{P}}}
\newcommand{\otP}{{\overline{\widetilde{P}}}}

\newcommand{\bq}{{\overline{q}}}
\newcommand{\bQ}{{\overline{Q}}}
\newcommand{\tr}{{\rm tr}}

\newcommand{\braket}[2]{\vev{#1 | #2}}
\newcommand{\vev}[1]{\left\langle #1 \right\rangle}
\title{Schr\"odinger functional formalism with domain-wall fermion}

\author{Yusuke Taniguchi\\
Institute of Physics, University of Tsukuba, Tsukuba, Ibaraki 305-8571,
Japan\\
E-mail: \email{tanigchi@het.ph.tsukuba.ac.jp}}

\received{\today}               %%
\accepted{\today}               %% These are for published papers.
\preprint{
UTHEP-520\\
UTCCS-P-20}

\abstract{
 Finite volume renormalization scheme is one of the most fascinating
 scheme for non-perturbative renormalization on lattice.
 By using the step scaling function one can follow running of
 renormalized quantities with reasonable cost.
 It has been established the Schr\"odinger functional is very
 convenient to define a field theory in a finite volume for the
 renormalization scheme.
 The Schr\"odinger functional, which is characterized by a
 Dirichlet boundary condition in temporal direction, is well defined and
 works well for the Yang-Mills theory and QCD with the Wilson fermion.
 However one easily runs into difficulties if one sets the same sort of
 the Dirichlet boundary condition for the overlap Dirac operator or the
 domain-wall fermion.
 In this paper we propose an orbifolding projection procedure to impose
 the Schr\"odinger functional Dirichlet boundary condition on the
 domain-wall fermion. 
}

\begin{document}

%\maketitle

%\newpage
\reseteqnum
%%%%%%%%%%%%%%%%%%%%%%%%%%%%%Section 1%%%%%%%%%%%%%%%%%%%%%%%%%%%%%%%
\section{Introduction}

Perturbative renormalization factor is a source of systematic errors in
numerical investigation of lattice QCD.
There has been progress in numerical simulation with dynamical fermions
nowadays and sources of systematic error is decreasing.
Evaluation of renormalization factors in non-perturbative method is
also required.
Finite volume renormalization scheme is one of the most fascinating
procedure to define non-perturbative renormalization scheme on lattice.
By using the step scaling function one can follow running of
renormalized quantities from low energy region to perturbative region
with reasonable cost for recent computers.
It has been established that the Schr\"odinger functional is very
convenient to define a field theory in a finite volume for
renormalization scheme.

The Schr\"odinger functional (SF) is defined as a transition amplitude
between two boundary states with finite time separation
\cite{Symanzik:1981wd,Luscher:1985iu,RossiTesta,Leroy:1990eh} 
\begin{eqnarray}
Z=\braket{C';x_0=T}{C;x_0=0}=\int{\cal D}\Phi e^{-S[\Phi]}
\end{eqnarray}
and is written in a path integral representation of the field theory
with some boundary condition.
For renormalization of a finite volume theory defined through the SF
the renormalization scale is introduced by a finite volume
$T\times L^3\sim L^4$ of the system.
The formulation is already accomplished for the non-linear
$\sigma$-model \cite{Luscher:1991wu}, the non-Abelian gauge theory
\cite{Luscher:1992an} and the QCD with the Wilson fermion
\cite{Sint:1993un,Sint:1995rb} including
${\cal O}(a)$ improvement procedure \cite{Luscher:1996sc,Luscher:1996vw}.
(See Ref.~\cite{Sint:2000vc} for review.)

Several renormalization quantities like running gauge coupling
\cite{Luscher:1992zx,Luscher:1993gh,deDivitiis:1994yz,Luscher:1995nr,Narayanan:1995ex,Sint:1995ch,DellaMorte:2004bc},
Z-factors and ${\cal O}(a)$ improvement factors
\cite{Jansen:1995ck,Luscher:1996ug,Luscher:1996jn,Jansen:1998mx,Capitani:1998mq,Ide:2004cd}
are extracted conveniently by using a Dirichlet boundary
conditions for spatial component of the gauge field
\begin{eqnarray}
 A_k(x)|_{x_0=0}=C_k(\vec{x}),\quad
 A_k(x)|_{x_0=T}=C_k'(\vec{x})
\label{eqn:DBCgauge}
\end{eqnarray}
and for the quark fields
\begin{eqnarray}
&&
P_+\psi(x)|_{x_0=0}=\rho(\vec{x}),\quad
P_-\psi(x)|_{x_0=T}=\rho'(\vec{x}),
\label{eqn:DBCpsi}
\\&&
\bpsi(x)P_-|_{x_0=0}=\brho(\vec{x}),\quad
\bpsi(x)P_+|_{x_0=T}=\brho'(\vec{x}),
\label{eqn:DBCbpsi}
\\&&
P_\pm=\frac{1\pm\gamma_0}{2}.
\end{eqnarray}
One of advantage of this Dirichlet boundary condition is that the system
acquire a mass gap proportional to $1/T$ and there is no infra-red
divergence.
The finite volume plays a role of an infra-red cut-off.
Field theory with Dirichlet boundary condition is shown to
be renormalizable for the pure gauge theory \cite{Luscher:1992an} and
QCD with the Wilson fermion \cite{Sint:1995rb}.

Although it is essential to adopt Dirichlet boundary condition for a
mass gap and renormalizability, it has a potential problem of zero mode
in fermion system.
For instance starting from a free Lagrangian
\begin{eqnarray}
{\cal L}=\bpsi\left(\gamma_\mu\p_\mu+m\right)\psi
\end{eqnarray}
with positive mass $m>0$ and the Dirichlet boundary condition
\begin{eqnarray}
P_-\psi|_{x_0=0}=0,\quad
P_+\psi|_{x_0=T}=0
\end{eqnarray}
the zero eigenvalue equation $\left(\gamma_0\p_0+m\right)\psi=0$ in
temporal direction allows a solution
\begin{eqnarray}
\psi=P_+e^{-mx_0}+P_-e^{-m(T-x_0)}
\end{eqnarray}
in $T\to\infty$ limit and a similar solution remains even for finite $T$
with an exponentially small eigenvalue $\propto e^{-mT}$.
In the SF formalism this solution is forbidden by adopting an ``opposite''
Dirichlet boundary condition \eqn{eqn:DBCpsi} and the system has a
finite gap even for $m=0$ \cite{Sint:1993un}.

For the Wilson fermion \cite{Sint:1993un} on lattice the Dirichlet boundary
condition is automatically chosen among
\begin{eqnarray}
P_\pm\psi|_{x_0=0}=0,\quad
P_\mp\psi|_{x_0=T}=0
\end{eqnarray}
depending on signature of the Wilson term.
For example if we adopt a typical signature of the Wilson term
\begin{eqnarray}
D_W=\gamma_\mu\frac{1}{2}\left(\nabla_\mu^*+\nabla_\mu\right)
-\frac{a}{2}\nabla_\mu^*\nabla_\mu+M
\label{eqn:DW}
\end{eqnarray}
the allowed Dirichlet boundary condition is the same as
\eqn{eqn:DBCpsi}.
In this case the zero mode solution is forbidden by choosing a
proper signature for the mass term; the mass should be kept positive
$M\ge0$ to eliminate the zero mode \cite{Sint:1993un}.

However as was discussed in the previous paper
\cite{Taniguchi:2004gf,Taniguchi:2005gh} 
this zero mode problem may become fatal in the overlap Dirac operator
\cite{Neuberger:1997fp,Neuberger:1998wv} and the domain-wall fermion
\cite{Kaplan:1992bt,Shamir:1993zy,Furman:1994ky}.
Both the overlap Dirac operator and the domain-wall fermion is defined
through the four dimensional Wilson Dirac operator \eqn{eqn:DW} but with
an opposite signature for the Wilson fermion mass parameter $M$
(domain-wall height) to the Wilson parameter $r$.
An opposite signature is necessary to impose heavy masses on the
doublers and a single massless mode to survive.
A requirement to the four dimensional Wilson Dirac operator is that
$D_W$ should not have a continuous zero mode.
If this is not the case the chiral Ward-Takahashi identity is broken
dynamically for the domain-wall fermion that the explicit breaking term
does not vanish \cite{Furman:1994ky}.
For the overlap Dirac operator a continuous zero mode may break locality
of the Dirac operator \cite{Hernandez:1998et}.

If the Dirichlet boundary condition \eqn{eqn:DBCpsi} \eqn{eqn:DBCbpsi}
is imposed to all fermion fields of the overlap Dirac operator or
the domain-wall fermion exponentially small eigenvalues are allowed in
the kernel $D_W$ because of an opposite signature of the Wilson
parameter and the domain-wall height.
Since these small eigenvalues are continuous in spatial momentum they
may be a lethal problem in large $T$ limit to break essential properties
of the chiral Dirac operator.

One may wonder that the small eigenvalues are boundary effect and should
be localized near the temporal boundary.
If one considers physics apart from the boundary there should be no
harm.
However this is not the case for our purpose to define renormalization
scheme.
In finite volume scheme the renormalization scale is given by a size of
the box, which is realized by considering a correlation function of
operators to be separated by an order of box size.
At least one of operators cannot be away from the boundary.
Furthermore it is convenient for the SF scheme to set one of the
operator at the boundary.

In order to solve this problem an orbifolding projection procedure was
proposed for the overlap Dirac operator in Ref.~\cite{Taniguchi:2004gf}.
\footnote{
After finishing this paper a new paper appeared to propose a method to
define chiral symmetric theory with the SF Dirichlet boundary condition
\cite{Luscher:2006df}.}
In this formulation we start from a theory on $S^1\times R^3$ and impose
orbifolding projection $S^1/Z_2$ on temporal direction.
Since we have set anti-periodic boundary condition in temporal direction
$S^1$ before projection we have a mass gap proportional to $1/T$, which
is not broken by the orbifolding.
Because of this mass gap we can avoid the zero mode problem of Dirichlet
boundary condition.

In this paper the orbifolding formulation of the SF boundary condition
is applied to the domain-wall fermion.
In section \ref{sec:dwf} the domain-wall fermion on $S^1\times T^3$ is
introduced.
Formulation of domain-wall fermion in finite volume with the SF boundary
condition is discussed in section \ref{sec:traditional}.
Application of orbifolding procedure to fermionic part is almost
straightforward as was discussed in Ref.~\cite{Taniguchi:2005gh}.
We can use the same kind of symmetry argument as in the previous paper
\cite{Taniguchi:2004gf}.
Difficulty is in a treatment of the Pauli-Villars field.
We adopted effective Dirac operator for this purpose.
The proper Dirichlet boundary condition \eqn{eqn:DBCpsi}
\eqn{eqn:DBCbpsi} may not be the unique choice to define a finite volume
renormalization scheme.
In section \ref{sec:twisted} a chirally twisted boundary condition is
discussed to define a finite volume field theory keeping a good property
of the SF boundary condition.
Section \ref{sec:conclusion} is devoted for conclusion.

\reseteqnum
%%%%%%%%%%%%%%%%%%%%%%%%%%%%%Section 3%%%%%%%%%%%%%%%%%%%%%%%%%%%%%%%
\section{Domain-wall fermion action}
\label{sec:dwf}

The purpose of this paper is to introduce the domain-wall fermion
system, with which we can define a finite volume renormalization scheme
(Schr\"odinger functional scheme).
The formulation for the pure Yang-Mills theory has been established in
Ref.~\cite{Luscher:1992an} by using a transition amplitude between two
boundary states (Schr\"odinger functional).
In this formulation the gauge field (link variable) lives in a finite
box $N_T\times N_L^3$ with a periodic boundary condition in spatial
direction and the SF Dirichlet boundary condition at the temporal
boundary
\begin{eqnarray}
&&
U_k(\vec{x},0)=W_k(\vec{x}),\quad
U_k(\vec{x},N_T)=W_k'(\vec{x}).
\label{eqn:sfgauge}
\end{eqnarray}
We shall adopt this procedure for the gauge part and treat the gauge
field as an external field in this paper.

The transition amplitude of the fermion field has been introduced for
the Wilson fermion using the transfer matrix in Ref.~\cite{Sint:1993un}.
The fermion field resides in the same finite box for the path integral
formalism with periodic or twisted boundary condition
\cite{Luscher:1996sc} in spatial direction and the SF Dirichlet boundary
condition \eqn{eqn:DBCpsi} and \eqn{eqn:DBCbpsi} in temporal direction.
This fermion system is renormalizable including a shift in the boundary
field $\rho$ and $\brho$ \cite{Sint:1995rb}.
Another specific property is that this system has a mass gap
proportional to the temporal length $1/T$ and the finite box serves as
an infra-red regulator.

We shall construct the domain-wall fermion system in a finite box
keeping the same sort of properties as the Wilson fermion;
(i) the theory has a mass gap proportional to $1/T$,
(ii) there are boundary fields $\rho$ and $\brho$ in temporal direction
and the theory is renormalizable including a shift in these fields.
If one naively impose the boundary condition \eqn{eqn:DBCpsi} and
\eqn{eqn:DBCbpsi} to all the fifth dimensional field $\psi(x,s)$ then
the chiral symmetry is broken ``dynamically'' as explained in the
introduction.
In order to avoid this problem we adopt an orbifolding procedure, where
we start from doubled time length $2N_T$ and fermion fields in the
finite box of length $N_T$ with the Dirichlet boundary condition is
realized by an orbifolding projection.
For this purpose we copy gauge configuration with the SF boundary
condition \eqn{eqn:sfgauge} into negative region and produce a time
reflection symmetric configuration, which satisfies
\begin{eqnarray}
&&
U_k(\vec{x},x_0)=U_k(\vec{x},-x_0),\quad
U_0(\vec{x},x_0)=U_0^\dagger(\vec{x},-x_0-1)
\label{eqn:extgauge}
\end{eqnarray}
as in the previous formulation of overlap Dirac operator
\cite{Taniguchi:2004gf}.
The periodic boundary condition is set with length $2N_T$
\begin{eqnarray}
U_\mu(\vec{x},x_0+2N_T)=U_\mu(\vec{x},x_0).
\end{eqnarray}

In this paper we adopt the Shamir's domain-wall fermion
\cite{Shamir:1993zy,Furman:1994ky} on a lattice
$2N_T\times N_L^3\times N_5$
\begin{eqnarray}
&&
S=\sum_{\vec{x},\vec{y}}\sum_{x_0,y_0=-N_T+1}^{N_T}\sum_{s,t=1}^{N_5}
\bpsi(x,s)D_{\rm dwf}(x,y;s,t)\psi(y,t).
\end{eqnarray}
%Vectors $\vec{x}$ and $\vec{y}$ represent the spatial coordinate,
$x_0$ and $y_0$ represent the temporal coordinate which runs
$-N_T+1\le x_0\le N_T$.
$s$ and $t$ are used for the fifth dimensional coordinate which runs
$1\le s\le N_5$.
For later use of orbifolding we set the anti-periodic boundary condition
in temporal direction
\begin{eqnarray}
&&
\psi(\vec{x},x_0+2N_T,s)=-\psi(\vec{x},x_0,s),\quad
\bpsi(\vec{x},x_0+2N_T,s)=-\bpsi(\vec{x},x_0,s).
\label{eqn:APB}
\end{eqnarray}

The Dirac operator is given as a five dimensional Wilson's one with
conventional Wilson parameter $r=1$ and negative mass parameter 
(domain-wall height) $-M$ with $0<M<2$
\begin{eqnarray}
D_{\rm dwf}(x,y;s,t)&=&
\gamma_MD_M-\frac{1}{2}D^2-M
\nn\\&=&
\left(
 \frac{-1+\gamma_0}{2}U_0(x)W^+_{x_0,y_0}
+\frac{-1-\gamma_0}{2}U_0^\dagger(y)W^-_{x_0,y_0}\right)
\delta_{x_i,y_i}\delta_{s,t}
\nn\\&&
+\left(
 \frac{-1+\gamma_i}{2}U_i(x)\delta_{y_i,x_i+1}
+\frac{-1-\gamma_i}{2}U_i^\dagger(y)\delta_{y_i,x_i-1}\right)
\delta_{x_0,y_0}\delta_{s,t}
\nn\\&&
+\left(
\frac{-1+\gamma_5}{2}\Omega^+(m_f)_{s,t}
+\frac{-1-\gamma_5}{2}\Omega^-(m_f)_{s,t}\right)
\delta_{x,y}
\nn\\&&
+(5-M)\delta_{x,y}\delta_{s,t},
\label{eqn:dirac-op}
\end{eqnarray}
where $W^\pm$ are hopping operator in temporal direction with
anti-periodic boundary condition, whose explicit form for $2N_T=6$ is
written as
\begin{eqnarray}
W^+_{x_0,y_0}=\pmatrix{
 0 & 1 & 0 & 0 & 0 & 0\cr
 0 & 0 & 1 & 0 & 0 & 0\cr
 0 & 0 & 0 & 1 & 0 & 0\cr
 0 & 0 & 0 & 0 & 1 & 0\cr
 0 & 0 & 0 & 0 & 0 & 1\cr
-1 & 0 & 0 & 0 & 0 & 0\cr},\quad
W^-=\left(W^+\right)^\dagger.
\end{eqnarray}
%where the index runs $x_0,y_0=-N_T+1,\cdots,N_T$.
$\Omega^\pm$ are hopping operator in fifth direction with Dirichlet
boundary condition (for massless case), whose matrix form for $N_5=6$
is given by
\begin{eqnarray}
\Omega^+(m_f)_{s,t}=\pmatrix{
0 & 1 & 0 & 0 & 0 & 0\cr
0 & 0 & 1 & 0 & 0 & 0\cr
0 & 0 & 0 & 1 & 0 & 0\cr
0 & 0 & 0 & 0 & 1 & 0\cr
0 & 0 & 0 & 0 & 0 & 1\cr
-m_f& 0 & 0 & 0 & 0 & 0\cr},\quad
\Omega^-(m_f)=\left(\Omega^+(m_f)\right)^\dagger.
\label{eqn:omega}
\end{eqnarray}
Here $m_f$ is a physical quark mass.

The physical quark field is defined by the fifth dimensional boundary
field with chiral projection
\begin{eqnarray}
&&
q(x)=\left(P_L\delta_{s,1}+P_R\delta_{s,N_5}\right)\psi(x,s),
\label{eqn:q}
\\&&
\bq(x)=\bpsi(x,s)\left(\delta_{s,N_5}P_L+\delta_{s,1}P_R\right),
\label{eqn:bq}
\\&&
P_{R/L}=\frac{1\pm\gamma_5}{2}.
\end{eqnarray}
The physical quark mass term is given as an ordinary form
${\cal L}_{\rm mass}=m_f\bq q$ with this quark field.

\reseteqnum
%%%%%%%%%%%%%%%%%%%%%%%%%%%%%Section 3%%%%%%%%%%%%%%%%%%%%%%%%%%%%%%%
\section{Schr\"odinger functional with conventional boundary condition}
\label{sec:traditional}

In this section we shall construct the domain-wall fermion system in
finite box, in which the conventional SF Dirichlet boundary condition
\eqn{eqn:DBCpsi} and \eqn{eqn:DBCbpsi} is satisfied by the physical
quark field.
This formulation will be done by making use of an orbifolding in
temporal direction.

\subsection{Orbifolding construction of SF boundary condition}
\label{sec:orbifolding}

Since we adopted (anti) periodic boundary condition in temporal
direction with period $2N_T$ fields live on $S^1$.
The orbifolding $S^1/Z_2$ is to identify the negative time coordinate
with the positive one $x_0=-x_0$.
Identification of fields on $S^1$ is performed according to the symmetry
of the theory including the time reflection.
A homogeneous Dirichlet boundary condition will appear at fixed points.
%depending the symmetry we adopt.

The time reversal symmetry of the domain-wall fermion is given by
\begin{eqnarray}
&&
\psi(\vec{x},x_0,s)\to\ovl{\Sigma}_{x_0,y_0;s,t}\psi(\vec{x},y_0,t),
\quad
\bpsi(\vec{x},x_0,s)\to\bpsi(\vec{x},y_0,t)\ovl{\Sigma}_{y_0,x_0;t,s},
\label{eqn:timeref}
\\&&
\ovl{\Sigma}_{x_0,y_0;s,t}=i\gamma_5\gamma_0R_{x_0,y_0}P_{s,t},
\end{eqnarray}
where $P$ is a parity transformation in fifth direction
$P_{s,t}\psi(\vec{x},x_0,t)=\psi(\vec{x},x_0,N_5-s+1)$, whose matrix
representation is
\begin{eqnarray}
{\cal P}_{s,t}=\pmatrix{
0 & 0 & 0 & 0 & 0 & 1 \cr
0 & 0 & 0 & 0 & 1 & 0 \cr
0 & 0 & 0 & 1 & 0 & 0 \cr
0 & 0 & 1 & 0 & 0 & 0 \cr
0 & 1 & 0 & 0 & 0 & 0 \cr
1 & 0 & 0 & 0 & 0 & 0 \cr},\quad
(N_5=6)
\end{eqnarray}
and $R$ is a time reflection operator acting on the temporal direction
$R_{x_0,y_0}\psi(\vec{x},y_0,s)=\psi(\vec{x},-x_0,s)$, whose matrix form
is given by
\begin{eqnarray}
R_{x_0,y_0}=\pmatrix{
0 & 0 & 0 & 0 & 1 & 0 \cr
0 & 0 & 0 & 1 & 0 & 0 \cr
0 & 0 & 1 & 0 & 0 & 0 \cr
0 & 1 & 0 & 0 & 0 & 0 \cr
1 & 0 & 0 & 0 & 0 & 0 \cr
0 & 0 & 0 & 0 & 0 &-1 \cr},\quad
(2N_T=6)
\end{eqnarray}
to satisfy anti-periodicity in $2N_T$.
We notice that $R$ has a symmetric fixed point $x_0=0$ and an
anti-symmetric fixed point $x_0=N_T$
\begin{eqnarray}
R\psi(\vec{x},0,s)=\psi(\vec{x},0,s),\quad
R\psi(\vec{x},N_T,s)=-\psi(\vec{x},N_T,s).
\end{eqnarray}
The domain-wall fermion Dirac operator is invariant under the time
reflection
\begin{eqnarray}
\left[\ovl{\Sigma},D_{\rm dwf}\right]=0
\label{eqn:timeref2}
\end{eqnarray}
since the reflection invariant gauge configuration \eqn{eqn:extgauge} is
adopted .

In order to realize the SF boundary condition at the fixed points we
need to combine the chiral transformation with the time reflection
\cite{Taniguchi:2004gf}.
The chiral transformation is given by a vector like rotation of fermion
field but with a different charge for two boundaries in fifth direction
\cite{Furman:1994ky} 
\begin{eqnarray}
\psi(x,s)\to iQ_{s,t}\psi(x,t),\quad
\bpsi(x,s)\to -\bpsi(x,t)iQ_{t,s},
\label{eqn:chiraltr}
\end{eqnarray}
where $Q$ is the vector charge matrix which flips sign in the middle of
the fifth direction
\begin{eqnarray}
Q_{s,t}=\pmatrix{
1 & 0 & 0 & 0 & 0 & 0\cr
0 & 1 & 0 & 0 & 0 & 0\cr
0 & 0 & 1 & 0 & 0 & 0\cr
0 & 0 & 0 &-1 & 0 & 0\cr
0 & 0 & 0 & 0 &-1 & 0\cr
0 & 0 & 0 & 0 & 0 &-1\cr},\quad
(N_5=6).
\label{eqn:Q}
\end{eqnarray}
We consider massless $m_f=0$ theory in this sub-section.

Here we should notice that this chiral rotation is not an exact symmetry
of the domain-wall fermion Dirac operator but we have an explicit
breaking term
\begin{eqnarray}
QD_{\rm dwf}Q-D_{\rm dwf}=2X,
\end{eqnarray}
where $X$ is a contribution from the middle layer, which picks up a
charge difference there
\begin{eqnarray}
X=\left(P_L\delta_{s,\frac{N_5}{2}}\delta_{t,\frac{N_5}{2}+1}
+P_R\delta_{s,\frac{N_5}{2}+1}\delta_{t,\frac{N_5}{2}}\right)\delta_{x,y}.
\end{eqnarray}
However it was discussed in Ref.~\cite{Furman:1994ky} that if we
consider correlation functions between the bilinear $\bpsi X\psi$
and the physical quark operators contribution is suppressed
exponentially in $N_5$ under the condition that the transfer matrix in
fifth direction has a gap from unity.
Furthermore the domain-wall fermion Dirac operator with explicit time
reflection invariance \eqn{eqn:timeref2} does not have index
\cite{Taniguchi:2004gf}, since the contribution to the index
\cite{Kikukawa:1999sy}
\begin{eqnarray}
\lim_{N_5\to\infty}a^4\sum_{x}\vev{\bpsi(x,s)\gamma_5X_{s,t}\psi(x,t)}
=-\lim_{N_5\to\infty}\tr\left(\gamma_5X\frac{1}{D_{\rm dwf}}\right)
\label{eqn:index}
\end{eqnarray}
can be shown to vanish by using anti-commutativity
$\left\{\gamma_5X,\ovl{\Sigma}\right\}=0$.
We expect that $X$ has no effect on anomaly.
We shall ignore this term in the following by constraining that we
treat the physical quark Green's functions only.

Another way to avoid the explicit breaking term is to include it into
the Dirac operator.
By using an anti-commutative nature $\left\{Q,X\right\}=0$ we can define
a chiral symmetric Dirac operator by
\begin{eqnarray}
D_{\rm dwf}^{\rm sym}=D_{\rm dwf}+X,
\label{eqn:symmetric}
\end{eqnarray}
which commutes with $Q$ exactly even at finite $N_5$.
The orbifolding projection in the following can be defined in an exact
sense.
A compensation of the exact chiral symmetry at finite $N_5$ is a
non-locality in the effective Dirac operator, which however is
suppressed exponentially in $N_5$.
Detailed property of this Dirac operator is deferred in appendix A.

Combining the time reversal transformation \eqn{eqn:timeref} and the
chiral transformation \eqn{eqn:chiraltr} we define the orbifolding
transformation
\begin{eqnarray}
&&
\psi(\vec{x},x_0,s)\to A_{x_0,y_0;s,t}\psi(\vec{x},y_0,t),\quad
\bpsi(\vec{x},x_0,s)\to\bpsi(\vec{x},y_0,t)A_{y_0,x_0;t,s},
\label{eqn:orbifoldingtr}
\\&&
A_{x_0,y_0;s,t}=\gamma_0\gamma_5(PQ)_{s,t}R_{x_0,y_0}.
\end{eqnarray}
The domain-wall fermion Dirac operator has time reversal symmetry
\eqn{eqn:timeref2} and we assume that the chiral transformation is an
exact symmetry of the Dirac operator
\begin{eqnarray}
\left[Q,D_{\rm dwf}\right]=0
\end{eqnarray}
by ignoring effect of the explicit breaking term $X$ or by adopting the
symmetric Dirac operator.
The orbifolding transformation becomes symmetry of the Dirac operator
\begin{eqnarray}
\left[A,D_{\rm dwf}\right]=0.
\label{eqn:symmetry}
\end{eqnarray}
In order to show this we may use a relation $\left\{P,Q\right\}=0$.

The operator $A$ satisfies a property $A^2=1$ and can be used to define a
projection operator.
The orbifolding identification of the fermion field is given by
projecting out the following symmetric sub-space
\begin{eqnarray}
\oP_-\psi(x,s)=0,\quad
\left(\bpsi\;\oP_-\right)(x,s)=0,\quad
\oP_\pm=\frac{1\pm{A}}{2}.
\label{eqn:orbifolding}
\end{eqnarray}
This projection relates fields in negative region to those in the
positive $\psi(\vec{x},-x_0,s)=\gamma_0\gamma_5PQ\psi(\vec{x},x_0,s)$,
which means fields in the negative is not independent.
As will be discussed in appendix B if we consider non-negative region
$0\le x_0\le T$, fields in the bulk $0<x_0<T$ is not constrained.
Only the boundary fields obey a projection condition
\begin{eqnarray}
&&
\ovl{P}_-\psi(\vec{x},0,s)=0,\quad
\ovl{P}_+\psi(\vec{x},N_T,s)=0
\label{eqn:projection-boundary1}
\\&&
\left(\bpsi\;\ovl{P}_-\right)(\vec{x},0,s)=0,\quad
\left(\bpsi\;\ovl{P}_+\right)(\vec{x},N_T,s)=0
\label{eqn:projection-boundary2}
\end{eqnarray}
with projection operator
\begin{eqnarray}
\ovl{P}_\pm=\frac{1\pm\ovl{\Gamma}}{2},\quad
\ovl{\Gamma}=\gamma_0\gamma_5PQ.
\label{eqn:projection}
\end{eqnarray}

The orbifolding projection for the physical quark field is given by
picking up the boundary components from the projected fermion field
\begin{eqnarray}
&&
\left(P_L\delta_{s,1}+P_R\delta_{s,N_5}\right)
\left(\oP_-\right)_{s,t}\psi(x,t)=\Pi_+q(x)=0,
\\&&
\bpsi(x,t)\left(\oP_-\right)_{t,s}
\left(\delta_{s,N_5}P_L+\delta_{s,1}P_R\right)=\bq(x)\Pi_-=0,
\\&&
\Pi_\pm=\frac{1\pm\Gamma}{2},\quad
\Gamma=\gamma_0R,
\label{eqn:four-dim-orb}
\end{eqnarray}
which turns out to be the same condition for the continuum theory in
Ref.~\cite{Taniguchi:2004gf}.
The proper homogeneous SF Dirichlet boundary condition is provided at
fixed points $x_0=0,N_T$ for the physical quark fields
\begin{eqnarray}
&&
P_+q(x)|_{x_0=0}=0,\quad
P_-q(x)|_{x_0=N_T}=0,
\label{eqn:DBC1}
\\&&
\bq(x)P_-|_{x_0=0}=0,\quad
\bq(x)P_+|_{x_0=N_T}=0.
\label{eqn:DBC2}
\end{eqnarray}
The massless orbifolded action is given by projection
\begin{eqnarray}
S_{\rm SF}=\sum\frac{1}{2}\bpsi D^{\rm SF}_{\rm dwf}\psi,\quad
D^{\rm SF}_{\rm dwf}=\oP_+D_{\rm dwf}\oP_+.
\label{eqn:massless}
\end{eqnarray}
We notice the massless SF Dirac operator $D^{\rm SF}_{\rm dwf}$ breaks
``chiral symmetry'' \eqn{eqn:chiraltr} explicitly by the projection
$\oP_+$.
However the symmetry breaking effect comes from the projection
\eqn{eqn:projection-boundary1} \eqn{eqn:projection-boundary2} at the
boundary.
Ordinary chiral Ward-Takahashi identity \cite{Furman:1994ky} is
satisfied in the bulk $0<x_0<N_T$ where fields are not constrained.

Since our orbifolded action is given by projecting onto a symmetric
sub-space of the theory and the orbifolding symmetry is not broken by
anomaly
\footnote{According to a similar discussion to that for chiral index
\eqn{eqn:index} we can easily show that orbifolding matrix $A$ does not
have index.}
the renormalizability is kept trivially
\footnote{Boundary source fields are introduced in later sub-section.}.

Our original theory on $S^1$ has a gap because of the anti-periodic
boundary condition.
This gap is kept intact after orbifolding, which can be confirmed
at tree level.
We have a Hermiticity relation for the SF Dirac operator
\begin{eqnarray}
\left(D^{\rm SF}_{\rm dwf}\right)^\dagger=
\gamma_5PD^{\rm SF}_{\rm dwf}\gamma_5P
\end{eqnarray}
and this Dirac operator connects the same Hilbert sub-space
\begin{eqnarray}
D^{\rm SF}_{\rm dwf} : \ovl{\cal H}_-\to\ovl{\cal H}_-,\quad
\ovl{\cal H}_-=\left\{\psi\left|\oP_-\psi=0\right.\right\}.
\label{eqn:subspace}
\end{eqnarray}
It is straightforward to solve the eigenvalue problem numerically at
tree level.
Here we omit the detail but we can easily see that the lowest eigenvalue
(a gap) converge to $\pi/2T$ in the continuum limit, which agrees with
that of continuum massless theory \cite{Sint:1993un}.

As will be discussed in appendix B the bulk part of this projected Dirac
operator is exactly the same as that of the ordinary domain-wall
fermion.
The physical quark fields satisfies the proper boundary condition.
Together with the renormalizability and existence of the mass gap
this orbifolded system is a strong candidate of QCD with the SF boundary
condition to define a finite volume scheme.
%Several properties are verified at tree level by using propagator in the
%next sub-section
%\footnote{A complete check of this expectation with gauge interaction
%will be given in later sections when we consider the effective action of
%the domain-wall fermion.}.

We have a comment on mass term.
We dropped quark mass term since it breaks the chiral symmetry.
However as was discussed in Ref.~\cite{Taniguchi:2004gf} it is possible
to introduce a mass term which is consistent with the orbifolding
symmetry \eqn{eqn:orbifoldingtr}.
One of candidates is
\begin{eqnarray}
S_{\rm mass}=\sum_{x} m_f\bq(x)\eta(x_0)q(x),
\label{eqn:mass}
\end{eqnarray}
where $\eta$ is an anti-symmetric step function
\begin{eqnarray}
&&
\eta(-x_0)=-\eta(x_0),\quad
\eta(x_0+2T)=\eta(x_0),
\nn\\&&
\eta(x_0)=1\quad{\rm for}\quad 0< x_0 <N_T.
\end{eqnarray}

\subsection{Free propagator}

In order to check that the orbifolded domain-wall fermion system
describes the QCD with the SF boundary condition properly we consider
the physical quark propagator at tree level.
The massless fermion propagator is given as an inverse of the projected
Dirac operator
\begin{eqnarray}
G^{\rm SF}_{\rm dwf}(x,y;s,t)
=2\left(D^{\rm SF}_{\rm dwf}\right)^{-1}_{x,y;s,t}
=2\left(\oP_+\frac{1}{D_{\rm dwf}}\oP_+\right)_{x,y;s,t}.
\label{eqn:prop}
\end{eqnarray}
where inverse is defined in the sub-space $\ovl{\cal H}_-$
\begin{eqnarray}
D^{\rm SF}_{\rm dwf}\left(D^{\rm SF}_{\rm dwf}\right)^{-1}
=\left(D^{\rm SF}_{\rm dwf}\right)^{-1}D^{\rm SF}_{\rm dwf}=\oP_+.
\end{eqnarray}

At tree level this propagator can be written in a simple form as
\begin{eqnarray}
G^{\rm SF}_{\rm dwf}(x,y;s,t)&=&
\frac{1}{N_L^3}\sum_{\vec{p}}e^{i\vec{p}\left(\vec{x}-\vec{y}\right)}
G^{\rm SF}_{\rm dwf}(\vec{p};x_0,y_0;s,t),
\\
G^{\rm SF}_{\rm dwf}(\vec{p};x_0,y_0;s,t)&=&
\frac{1}{2aN_T}\sum_{n=-N_T+1}^{N_T}
\left(\frac{1}{D_{\rm dwf}(p)}\right)_{s,t'}
\biggl\{\left(e^{ip_0(x_0-y_0)}+e^{ip_0(x_0+y_0)}\right)
\left(\ovl{P}_+\right)_{t',t}
\nn\\&&\quad
+\left(e^{ip_0(x_0-y_0)}-e^{ip_0(x_0+y_0)}\right)\left(\ovl{P}_-\right)_{t',t}
\biggr\},
\end{eqnarray}
where the projection operator $\ovl{P}_\pm$ is defined in
\eqn{eqn:projection}.
The temporal momentum $p_0$ satisfies the quantization condition
\begin{eqnarray}
p_0=\frac{2n-1}{2N_T}\pi,\quad
-N_T+1\le n \le N_T
\label{eqn:p0}
\end{eqnarray}
for anti-periodicity in $2N_T$.
$D_{\rm dwf}(p)$ is the domain-wall fermion Dirac operator in momentum
space without orbifolding projection
\begin{eqnarray}
&&
aD_{\rm dwf}(p)=i\gamma_\mu\sin ap_\mu+W(p)-P_L\Omega^+-P_R\Omega^-,
\\&&
W=1-M+\sum_{\mu}\left(1-\cos ap_\mu\right).
\label{eqn:W}
\end{eqnarray}
The explicit form of its inverse can be derived according to
Ref.~\cite{Shamir:1993zy}, which we defer to appendix C.

The physical quark propagator is given by selecting the contribution
from the boundary fields in fifth direction
\begin{eqnarray}
G^{\rm SF}_{\rm quark}(x,y)&=&
%2\left(P_L\delta_{s,1}+P_R\delta_{s,N_5}\right)
%\left(\oP_+\frac{1}{D_{\rm dwf}}\oP_+\right)_{x,y;s,t}
%\left(\delta_{t,N_5}P_L+\delta_{t,1}P_R\right)
\left(P_L\delta_{s,1}+P_R\delta_{s,N_5}\right)
G^{\rm SF}_{\rm dwf}(x,y;s,t)
\left(\delta_{t,N_5}P_L+\delta_{t,1}P_R\right)
\nn\\&=&
2\left(\Pi_-G_{\rm quark}\Pi_+\right)_{x,y},
\label{eqn:qprop}
\end{eqnarray}
where
\begin{eqnarray}
G_{\rm quark}(x,y)=
\left(P_L\delta_{s,1}+P_R\delta_{s,N_5}\right)
\left(\frac{1}{D_{\rm dwf}}\right)_{x,y;s,t}
\left(\delta_{t,N_5}P_L+\delta_{t,1}P_R\right)
\end{eqnarray}
is the physical quark propagator in $2N_T\times N_L^3$ space-time
without any projection.
The proper Dirichlet boundary conditions \cite{Luscher:1996vw}
\begin{eqnarray}
&&
P_+G^{\rm SF}_{\rm quark}(x,y)|_{x_0=0}=0,\quad
P_-G^{\rm SF}_{\rm quark}(x,y)|_{x_0=N_T}=0,
\\&&
G^{\rm SF}_{\rm quark}(x,y)|_{y_0=0}P_-=0,\quad
G^{\rm SF}_{\rm quark}(x,y)|_{y_0=N_T}P_+=0
\end{eqnarray}
are satisfied for this quark propagator because of the projection
$\Pi_\pm$.
By ignoring sub-leading terms in $e^{-N_5}$ the propagator takes the
following form at tree level
\begin{eqnarray}
a^3\sum_{\vec{x}}e^{-i\vec{p}\left(\vec{x}-\vec{y}\right)}
G^{\rm SF}_{\rm quark}(x,y)&=&
\frac{1}{2aN_T}\sum_{n=-N_T+1}^{N_T}
\left(\frac{i\gamma_\mu \sin p_\mu}{1-e^{\alpha}W(p)}\right)
e^{ip_0x_0}
\nn\\&\times&
\left\{\left(e^{-ip_0y_0}+e^{ip_0y_0}\right)P_+
+\left(e^{-ip_0y_0}-e^{ip_0y_0}\right)P_-\right\},
\end{eqnarray}
which can be shown to approach to the continuum SF propagator of
Ref.~\cite{Luscher:1996vw} without any ${\cal O}(a)$ term.
% except for a constant normalization factor.
This system has no extra zero mode we encountered in naive formulation
and we conclude that this is equivalent to the QCD with SF boundary
condition.

\subsection{Surface term}

In the orbifolding construction of the SF formalism only the homogeneous
boundary condition \eqn{eqn:DBC1} \eqn{eqn:DBC2} can be introduced.
However in general SF formalism the Dirichlet boundary condition is
inhomogeneous as \eqn{eqn:DBCpsi} and \eqn{eqn:DBCbpsi}.
The boundary values $\rho, \cdots, \ovl{\rho}'$ are regarded as external
source fields coupled to the dynamical fields and the correlation
functions involving the boundary fields
\begin{eqnarray}
&&
\zeta(\vec{x})=\frac{\delta}{\delta\ovl{\rho}(\vec{x})},\quad
\ovl{\zeta}(\vec{x})=-\frac{\delta}{\delta\rho(\vec{x})},
\\&&
\zeta'(\vec{x})=\frac{\delta}{\delta\ovl{\rho}'(\vec{x})},\quad
\ovl{\zeta}'(\vec{x})=-\frac{\delta}{\delta\rho'(\vec{x})}
\end{eqnarray}
are used conveniently to extract the renormalization factors.

Coupling between the boundary source fields and the dynamical
fields are not introduced automatically in our formulation since the
boundary value vanishes by projection.
In other words our formulation is protected by an orbifolding symmetry.
The action \eqn{eqn:massless} is invariant under a transformation
\begin{eqnarray}
&&
\delta\left(\oP_+\psi\right)(x,s)=\alpha\left(\oP_+\psi\right)(x,s),
\label{eqn:inftr1}
\\&&
\delta\left(\bpsi\;\oP_+\right)(x,s)=-\alpha\left(\bpsi\;\oP_+\right)(x,s),
\label{eqn:inftr2}
\end{eqnarray}
where remaining degrees of freedom $\oP_-\psi$ and $\bpsi\;\oP_-$ are kept
intact.
The boundary source fields are elements of $\oP_-\psi$ and
$\bpsi\;\oP_-$.
Since the surface term is given to connect the boundary source fields
and the dynamical fields $\oP_+\psi$, $\bpsi\;\oP_+$ it is not consistent
with the orbifolding symmetry.

In this paper we define a surface term as an orbifolding symmetry
breaking term, which is consistent with other symmetries of the
orbifolded domain-wall fermion.
The symmetries are;
parity
\begin{eqnarray}
\psi(x,s)\to\gamma_0P_{st}\psi(-\vec{x},x_0,t),\quad
\bpsi(x,s)\to\bpsi(-\vec{x},x_0,t)\gamma_0P_{ts},
\end{eqnarray}
charge conjugation
\begin{eqnarray}
&&
\psi(x,s)\to CP_{st}\bpsi^T(x,t),\quad
\bpsi(x,s)\to \psi^T(x,t)P_{ts}\left(-C^{-1}\right),\quad
C=\gamma_2\gamma_0
\end{eqnarray}
and chiral symmetry \eqn{eqn:chiraltr} in the bulk $0<x_0<N_T$, where
chiral Ward-Takahashi identity of Ref.~\cite{Furman:1994ky} is satisfied.

Using the orbifolding projection \eqn{eqn:orbifolding} it is easy to
show that the orbifolding transformation \eqn{eqn:inftr1}
\eqn{eqn:inftr2} is a ``chiral'' transformation only at the boundary
\begin{eqnarray}
&&
\delta\left(\ovl{P}_+\psi\right)(\vec{x},0,s)
=\alpha\left(\ovl{P}_+\psi\right)(\vec{x},0,s),
\\&&
\delta\left(\bpsi\;\ovl{P}_+\right)(\vec{x},0,s)
=-\alpha\left(\bpsi\;\ovl{P}_+\right)(\vec{x},0,s),
\\&&
\delta\left(\ovl{P}_-\psi\right)(\vec{x},N_T,s)
=\alpha\left(\ovl{P}_-\psi\right)(\vec{x},N_T,s),
\\&&
\delta\left(\bpsi\;\ovl{P}_-\right)(\vec{x},N_T,s)
=-\alpha\left(\bpsi\;\ovl{P}_-\right)(\vec{x},N_T,s)
\end{eqnarray}
and is a vector $U(1)$ transformation in the bulk $0<x_0<N_T$
\begin{eqnarray}
\delta\psi(\vec{x},x_0,s)=\alpha\psi(\vec{x},x_0,s),\quad
\delta\bpsi(\vec{x},x_0,s)=-\alpha\bpsi(\vec{x},x_0,s).
\end{eqnarray}
This is a symmetry of the folded theory discussed in appendix B.
It is obviously inappropriate to break this vector $U(1)$ symmetry in
the bulk.
The ``chiral'' symmetry at the boundary can be broken by a physical quark
mass term $m_f\bq q$ keeping the bulk vector symmetry.
But this is forbidden by the chiral Ward-Takahashi identity in the bulk.
The symmetry should be broken only at the boundary.

We introduce boundary source fields as a component of projected out
degrees of freedom in \eqn{eqn:projection-boundary1} and
\eqn{eqn:projection-boundary2}
\begin{eqnarray}
&&
\lambda(\vec{x},s)=\ovl{P}_-\psi(\vec{x},0,s),\quad
\lambda'(\vec{x},s)=\ovl{P}_+\psi(\vec{x},N_T,s),
\\&&
\blambda(\vec{x},s)=\left(\bpsi\;\ovl{P}_-\right)(\vec{x},0,s),\quad
\blambda'(\vec{x},s)=\left(\bpsi\;\ovl{P}_+\right)(\vec{x},N_T,s)
\end{eqnarray}
The orbifolding symmetry breaking term takes the form
\begin{eqnarray}
S_{\rm breaking}&=&
\blambda(\vec{x},s)\hat{O}_{st}\ovl{P}_+\psi(\vec{x},0,t)
+\left(\bpsi\;\ovl{P}_+\right)(\vec{x},0,s)\hat{O}_{st}\lambda(\vec{x},t)
\nn\\&&
+\blambda'(\vec{x},s)\hat{O}_{st}\ovl{P}_-\psi(\vec{x},N_T,t)
+\left(\bpsi\;\ovl{P}_-\right)(\vec{x},N_T,s)\hat{O}_{st}\lambda'(\vec{x},t),
\end{eqnarray}
where $\hat{O}$ is a local operator which anti-commute with
$\ovl{\Gamma}=\gamma_0\gamma_5PQ$.
Candidates of $\hat{O}$ are $\gamma_0$, $\gamma_5$, $Q$, $P$ and
\begin{eqnarray}
&&
K(u)_{st}=\left(P_L\delta_{s,N_5+1-u}+P_R\delta_{s,u}\right)
\left(P_L\delta_{t,u}+P_R\delta_{t,N_5+1-u}\right),
\label{eqn:K}
\\&&
\wt{K}(u)_{st}=\left(P_R\delta_{s,N_5+1-u}+P_L\delta_{s,u}\right)
\left(P_R\delta_{t,u}+P_L\delta_{t,N_5+1-u}\right),
\label{eqn:wtK}
\end{eqnarray}
where $K(1)$ produces a physical quark mass term
$\bpsi K(1)\psi=\bq q$.

Among these candidates $\gamma_5$ and $Q$ are forbidden by the parity
symmetry.
$\gamma_0$ is not consistent with the charge conjugation.
$P$, $K(u)$ and $\wt{K}(u)$ break chiral symmetry, which however is
broken at the boundary.
Since $P$, $K(u)$ and $\wt{K}(u)$ are consistent with parity and charge
conjugation they are proper candidates of orbifolding symmetry breaking
term.
However we adopt only $K(1)$ in this paper to reproduce the same
surface term as the continuum theory.
We define the surface term as
\begin{eqnarray}
S_{\rm surface}&=&
-a^3\sum_{\vec{x}}\Bigl(
\blambda(\vec{x},s)K(1)_{st}\ovl{P}_+\psi(\vec{x},0,t)
+\left(\bpsi\;\ovl{P}_+\right)(\vec{x},0,s)K(1)_{st}\lambda(\vec{x},t)
\nn\\&&
+\blambda'(\vec{x},s)K(1)_{st}\ovl{P}_-\psi(\vec{x},N_T,t)
+\left(\bpsi\;\ovl{P}_-\right)(\vec{x},N_T,s)K(1)_{st}\lambda'(\vec{x},t)
\Bigr)
\nn\\&=&
a^3\sum_{\vec{x}}\Bigl(
-\left.\ovl{\rho}(\vec{x})P_-q(x)\right|_{x_0=0}
-\left.\bq(x)P_+\rho(\vec{x})\right|_{x_0=0}
\nn\\&&
-\left.\ovl{\rho}'(\vec{x})P_+q(x)\right|_{x_0=N_T}
-\left.\bq(x)P_-\rho'(\vec{x})\right|_{x_0=N_T}
\Bigr),
\label{eqn:surface}
\end{eqnarray}
where $q$ and $\bq$ are active dynamical fields at the temporal
boundary.
$\rho$ and $\brho$ are boundary source fields for the physical quark
fields
\begin{eqnarray}
&&
P_+q(x)|_{x_0=0}=\rho(\vec{x}),\quad
P_-q(x)|_{x_0=N_T}=\rho'(\vec{x}),
\label{eqn:DBCq}
\\&&
\bq(x)P_-|_{x_0=0}=\brho(\vec{x}),\quad
\bq(x)P_+|_{x_0=N_T}=\brho'(\vec{x}).
\label{eqn:DBCbq}
\end{eqnarray}

Since this surface term is not a general orbifolding symmetry breaking
term, general anticipation is that we will need all sort of breaking
terms to renormalize quantum corrections.
However if we consider Green functions constructed with physical quark
operators only we may expect that quantum corrections which appear in
these Green functions can be renormalized into a shift of physical
operators and physical quark source fields $\rho$, $\brho$, $\rho'$ and
$\brho'$.
The situation is similar to the physical quark mass term $m_f\bq q$.
This term is not a general chiral symmetry breaking term.
However if we consider a Green function of the physical quark fields
only, quantum corrections which appear with chiral symmetry breaking
can be renormalized into the physical operators and mass term.
We did not need all the breaking terms for renormalization if we consider
the physical quark Green function.
Explicit calculation is necessary to confirm this expectation for source
fields.

We check validity of this surface term at tree level.
According to Ref.~\cite{Luscher:1996vw} we introduce the generating
functional
\begin{eqnarray}
Z_F\left[\ovl{\rho}',\rho';\ovl{\rho},\rho;\ovl{\eta},\eta;U\right]&=&
\int D\psi D\bpsi \exp\biggl\{
-S_F\left[U,\bpsi,\psi;\ovl{\rho}',\rho',\ovl{\rho},\rho\right]
\nn\\&&
+a^4\sum_{x,s}\left(\bpsi(x,s)\eta(x,s)+\ovl{\eta}(x,s)\psi(x,s)\right)
\biggr\},
\end{eqnarray}
where $\eta(x)$ and $\ovl{\eta}(x)$ are source fields for the fermion
fields and the total action $S_F$ is given as a sum of the bulk action
\eqn{eqn:massless} and the surface term \eqn{eqn:surface}.
We notice that the fermion fields $\psi$ and $\bpsi$ obey the
orbifolding condition \eqn{eqn:orbifolding}.
The correlation functions between the boundary fields are derived with
the same procedure as Ref.~\cite{Luscher:1996vw}.
\begin{eqnarray}
&&
\vev{\psi(x,s)\bpsi(y,t)}=G^{\rm SF}_{\rm dwf}(x,y;s,t),
\\&&
\vev{q(x)\bq(y)}=G^{\rm SF}_{\rm quark}(x,y),
\\&&
\vev{q(x)\ovl{\zeta}(\vec{y})}=
\left.G^{\rm SF}_{\rm quark}(x,y)P_+\right|_{y_0=0}
\\&&
\vev{q(x)\ovl{\zeta}'(\vec{y})}=
\left.G^{\rm SF}_{\rm quark}(x,y)P_-\right|_{y_0=N_T},
\\&&
\vev{\zeta(\vec{x})\bq(y)}=P_-\left.G^{\rm SF}_{\rm quark}(x,y)\right|_{x_0=0},
\\&&
\vev{\zeta'(\vec{x})\bq(y)}=P_+\left.G^{\rm SF}_{\rm quark}(x,y)
\right|_{x_0=N_T},
\\&&
\vev{\zeta(\vec{x})\ovl{\zeta}(\vec{y})}=
P_-\left.G^{\rm SF}_{\rm quark}(x,y)P_+\right|_{x_0=0,y_0=0},
\\&&
\vev{\zeta(\vec{x})\ovl{\zeta}'(\vec{y})}=
P_-\left.G^{\rm SF}_{\rm quark}(x,y)P_-\right|_{x_0=0,y_0=N_T},
\\&&
\vev{\zeta'(\vec{x})\ovl{\zeta}(\vec{y})}=
P_+\left.G^{\rm SF}_{\rm quark}(x,y)P_+\right|_{x_0=N_T,y_0=0},
\\&&
\vev{\zeta'(\vec{x})\ovl{\zeta}'(\vec{y})}=
P_+\left.G^{\rm SF}_{\rm quark}(x,y)P_-\right|_{x_0=N_T,y_0=N_T}.
\end{eqnarray}
The propagator $G^{\rm SF}_{\rm dwf}$ and $G^{\rm SF}_{\rm quark}$ are
given in \eqn{eqn:prop} and \eqn{eqn:qprop}.
We notice that the above propagators between the boundary fields and
physical quark fields approach to the continuum SF boundary propagator
without any ${\cal O}(a)$ term at tree level.

\subsection{Effective action of the domain-wall fermion}

In order to perform numerical simulation with dynamical fermion we need
to introduce the Pauli-Villars field to cancel bulk contribution in
fifth direction.
The Pauli-Villars field is a four component complex scalar and its
action is given by
\begin{eqnarray}
S_{\rm PV}=\sum_{\vec{x},\vec{y}}\sum_{x_0,y_0=-N_T+1}^{N_T}\sum_{s,t=1}^{N_5}
\bphi(x,s)D_{\rm PV}(x,y;s,t)\phi(y,t),
\end{eqnarray}
where Dirac operator for the Pauli-Villars field is given in the same
form as the domain-wall fermion Dirac operator \eqn{eqn:dirac-op} with
$m_f=1$ \begin{eqnarray}
D_{\rm PV}=D_{\rm dwf}(m_f=1).
\end{eqnarray}
This Dirac operator does not commute with the orbifolding operator
$A=\gamma_0\gamma_5PQR$ because of the mass term.
It is not straightforward to introduce the Pauli-Villars field by
orbifolding.
In this paper we propose to implement it by the effective Dirac operator
\cite{Kikukawa:1999sy,Neuberger:1997bg}.

The effective Dirac operator appears in an effective action of the
physical quark field \eqn{eqn:q} \eqn{eqn:bq} and ``physical''
Pauli-Villars field
\begin{eqnarray}
&&
Q(x)=\left(P_L\delta_{s,1}+P_R\delta_{s,N_5}\right)\phi(x,s),
\label{eqn:physical-PV1}
\\&&
\bQ(x)=\bphi(x,s)\left(\delta_{s,N_5}P_L+\delta_{s,1}P_R\right).
\label{eqn:physical-PV2}
\end{eqnarray}
The effective action is given by integrating out all the bulk fields
other than physical fields at the fifth dimensional boundary
\cite{Kikukawa:1999sy}
\begin{eqnarray}
S_{\rm eff}=\bq(x)\left(D_{\rm eff}\right)_{xy}q(y)
+\bQ(x)\left(D_{\rm eff}+1\right)_{xy}Q(y).
\end{eqnarray}
In its derivation the effective Dirac operator $D_{\rm eff}$ is given
as an inverse of the full physical quark propagator
\begin{eqnarray}
D_{\rm eff}=\frac{1}{\vev{q\bq}},
\end{eqnarray}
whose explicit form is
\begin{eqnarray}
D_{\rm eff}=\frac{1+\gamma_5S}{1-\gamma_5S},\quad
S=\frac{1-T^{N_5}}{1+T^{N_5}},\quad
T=\frac{1-H'}{1+H'},\quad
H'=\gamma_5D_W\frac{1}{2+D_W}.
\label{eqn:truncated}
\end{eqnarray}
Here $D_W$ is a four dimensional Wilson Dirac operator with negative
mass $-2<-M<0$.
In $N_5\to\infty$ limit the Dirac operator becomes
\begin{eqnarray}
D_{\rm eff}=\frac{1+\gamma_5\epsilon\left(\wt{H}\right)}
{1-\gamma_5\epsilon\left(\wt{H}\right)},\quad
T=e^{-\wt{H}},
\label{eqn:effective}
\end{eqnarray}
where $\epsilon(x)$ is a sign function
\begin{eqnarray}
\epsilon(x)=\frac{x}{\sqrt{x^2}}.
\end{eqnarray}
We can easily check that this Dirac operator is exactly chiral symmetric
\begin{eqnarray}
\left\{\gamma_5,D_{\rm eff}\right\}=0
\end{eqnarray}
in $N_5\to\infty$ limit and should be non-local to satisfy the
Nielsen-Ninomiya's no-go theorem.

The effective Dirac operator is related to the original domain-wall
fermion and the Pauli-Villars Dirac operator through determinant
\begin{eqnarray}
\det\frac{1}{D_{\rm PV}}D_{\rm dwf}=\det\frac{D_{\rm eff}}{D_{\rm eff}+1}
=\det D_{N_5},
\end{eqnarray}
where $D_{N_5}$ is a truncated overlap Dirac operator
\cite{Kikukawa:1999sy,Neuberger:1997bg}.
Hereafter we take $N_5\to\infty$ limit implicitly and write
$D_{N_5\to\infty}=D_{\rm OD}$.
In terms of the domain-wall fermion the overlap Dirac operator is
defined as
\begin{eqnarray}
D_{\rm OD}=\frac{D_{\rm eff}}{D_{\rm eff}+1}.
\end{eqnarray}
$D_{\rm OD}$ satisfies the Ginsparg-Wilson relation
\cite{Ginsparg:1981bj}
\begin{eqnarray}
\left\{\gamma_5,D_{\rm OD}\right\}=2D_{\rm OD}\gamma_5D_{\rm OD}.
\end{eqnarray}
If we introduce physical quark mass term we have a massive overlap Dirac
operator through determinant
\begin{eqnarray}
D_{\rm OD}(m_f)=\frac{D_{\rm eff}+m_f}{D_{\rm eff}+1}
=D_{\rm OD}+m_f\left(1-D_{\rm OD}\right).
\label{eqn:massiveOD}
\end{eqnarray}

The effective Dirac operator of the orbifolded domain-wall fermion
system is defined in a similar way.
Since the four dimensional Wilson Dirac operator $D_W$ commute with the
four dimensional time reflection operator $\Sigma=i\gamma_5\gamma_0R$ we
have following anti-commutation relations
\begin{eqnarray}
\left\{\Sigma,H'\right\}=0,\quad
\left\{\Sigma,\wt{H}\right\}=0.
\end{eqnarray}
By using these relations we can easily show that the effective Dirac
operator \eqn{eqn:effective} anti-commute with the four dimensional
orbifolding operator $\Gamma$ defined in \eqn{eqn:four-dim-orb}
\begin{eqnarray}
\left\{\Gamma, D_{\rm eff}\right\}=0.
\end{eqnarray}
The massless overlap Dirac operator defined in the above satisfy
``Ginsparg-Wilson relation'' for the orbifolding transformation
\cite{Taniguchi:2004gf}
\begin{eqnarray}
\left\{\Gamma,D_{\rm OD}\right\}=2D_{\rm OD}\Gamma D_{\rm OD}.
\label{eqn:GW-relation}
\end{eqnarray}

We define the Schr\"odinger functional effective Dirac operator as an
inverse of the orbifolded full quark propagator \eqn{eqn:qprop}
\begin{eqnarray}
D_{\rm eff}^{\rm SF}=\Pi_+\frac{1}{\vev{q\bq}}\Pi_-=\Pi_+D_{\rm eff}\Pi_-,
\end{eqnarray}
where inverse means that in a sub-space
\begin{eqnarray}
D_{\rm eff}^{\rm SF}G^{\rm SF}_{\rm quark}=2\Pi_+,\quad
G^{\rm SF}_{\rm quark}D_{\rm eff}^{\rm SF}=2\Pi_-.
\end{eqnarray}
Contribution from the Pauli-Villars field is introduced to reproduce the
Schr\"odinger functional overlap Dirac operator defined in
Ref.~\cite{Taniguchi:2004gf}
\footnote{
The Ginsparg-Wilson relation \eqn{eqn:GW-relation} in this paper is
different from that in Ref.~\cite{Taniguchi:2004gf} by factor two and so
is the definition of $\wh{\Gamma}$.}
\begin{eqnarray}
D_{\rm OD}^{\rm SF}=\Pi_+D_{\rm eff}\Pi_-\frac{1}{D_{\rm eff}+1}
=\Pi_+D_{\rm OD}\hP_-,
\end{eqnarray}
where
\begin{eqnarray}
\hP_\pm=\frac{1\pm\wh{\Gamma}}{2},\quad
\wh{\Gamma}=\Gamma\left(1-2D_{\rm OD}\right).
\end{eqnarray}

This is not a unique definition of the SF overlap Dirac operator but we
can define another Dirac operator as
\begin{eqnarray}
\ovl{D}_{\rm OD}^{\rm SF}=\frac{1}{D_{\rm eff}+1}\Pi_+D_{\rm eff}\Pi_-.
\end{eqnarray}
These two Dirac operators are related by unitary operators
\begin{eqnarray}
u=\frac{1+\Sigma}{2}\left(1-2D_{\rm OD}\right)+\frac{1-\Sigma}{2},\quad
u'=\gamma_5u\gamma_5
\end{eqnarray}
by
\begin{eqnarray}
uD_{\rm OD}^{\rm SF}u^\dagger=\ovl{D}_{\rm OD}^{\rm SF},\quad
u'^\dagger D_{\rm OD}^{\rm SF}u'=\ovl{D}_{\rm OD}^{\rm SF}.
\end{eqnarray}
Here we used a fact that the effective and the overlap Dirac operators
commute with the four dimensional time reflection operator $\Sigma$.

As was discussed in Ref.~\cite{Taniguchi:2004gf} the SF overlap Dirac
operator does not have $\gamma_5$ Hermiticity relation.
Instead we have
\begin{eqnarray}
\left(D^{\rm SF}_{\rm OD}\right)^\dagger
=\gamma_5\ovl{D}_{\rm OD}^{\rm SF}\gamma_5.
\end{eqnarray}
In order to define real fermion determinant we may need even numbers of
flavours and different Dirac operators for each flavours.
An example for two flavours case is
\begin{eqnarray}
D_{\rm SF}^{(2)}&=&
\pmatrix{D^{\rm SF}_{\rm OD}\cr &\ovl{D}_{\rm OD}^{\rm SF}\cr}.
\end{eqnarray}
We notice that $U(2)$ vector flavour symmetry is broken to
$U(1)\times U(1)$.
Determinant of this Dirac operator is
\begin{eqnarray}
\det D_{\rm SF}^{(2)}
=\det D_{\rm SF}^{(2)}\gamma_5
=\det_{{\cal H}_-}\left(
\Pi_+D_{\rm eff}\frac{1}{D_{\rm eff}+1}
\frac{1}{D_{\rm eff}^\dagger+1}D_{\rm eff}^\dagger\Pi_+\right),
\end{eqnarray}
which is re-written in terms of pseudo-fermion field $\chi$
\begin{eqnarray}
\det D_{\rm SF}^{(2)}=
\int{\cal D}\left(\Pi_+\chi^\dagger\right){\cal D}\left(\Pi_+\chi\right)
\exp\left(-\chi^\dagger
\Pi_+\left(\frac{1}{D_{\rm eff}^\dagger}+1\right)
\left(\frac{1}{D_{\rm eff}}+1\right)\Pi_+\chi\right).
\end{eqnarray}
The determinant is defined in a sub-space
${\cal H}_-=\left\{\psi|\Pi_-\psi=0\right\}$ of eigenfunctions.
In evaluation of the fermion force we need to calculate
\begin{eqnarray}
\left(\frac{1}{D_{\rm eff}}+1\right)^{-1}
=\left(\vev{q\bq}+1\right)^{-1},
\end{eqnarray}
which corresponds to inverse of the overlap Dirac operator.

The orbifolded effective Dirac operator is modified as follows when we
introduce the mass term \eqn{eqn:mass}
\begin{eqnarray}
D_{\rm eff}^{\rm SF}(m_f)=\frac{1}{2}\Pi_+\left(D_{\rm eff}+m_f\eta\right)
\Pi_-.
\end{eqnarray}
Taking into account a contribution from the Pauli-Villars Dirac operator
the massive SF overlap Dirac operator is defined as
\begin{eqnarray}
D_{\rm OD}^{\rm SF}(m_f)&=&
\frac{1}{2}\Pi_+\left(D_{\rm eff}+m_f\eta\right)\Pi_-\frac{1}{D_{\rm eff}+1}
=\frac{1}{2}\Pi_+\left(D_{\rm OD}+m_f\eta\left(1-D_{\rm OD}\right)\right)\hP_-,
\nn\\
\\
\ovl{D}_{\rm OD}^{\rm SF}(m_f)&=&
\frac{1}{2}\frac{1}{D_{\rm eff}+1}\Pi_+\left(D_{\rm eff}+m_f\eta\right)\Pi_-.
\end{eqnarray}
Although we do not have a unitary transformation to relate
$D_{\rm OD}^{\rm SF}(m_f)$ and $\ovl{D}_{\rm OD}^{\rm SF}(m_f)$ we have a
Hermiticity relation
\begin{eqnarray}
\left(D^{\rm SF}_{\rm OD}(m_f)\right)^\dagger
=\gamma_5\ovl{D}_{\rm OD}^{\rm SF}(m_f)\gamma_5.
\end{eqnarray}
We also need even numbers of flavours to define a real fermion
determinant.

\reseteqnum
%%%%%%%%%%%%%%%%%%%%%%%%%%%%%Section 3%%%%%%%%%%%%%%%%%%%%%%%%%%%%%%%
\section{Schr\"odinger functional with twisted boundary condition}
\label{sec:twisted}

In the previous section we presented an orbifolding formulation of
domain-wall fermion in finite box, in which the homogeneous proper
boundary condition \eqn{eqn:DBC1} \eqn{eqn:DBC2} is satisfied.
This is a solution of our purpose to define a finite volume
renormalization scheme.
However this may not be the unique solution of our requirement that the
theory has a mass gap and is kept to be renormalizable in a finite box.
In this section we propose another orbifolding formulation to adopt
chirally twisted boundary condition
\cite{Taniguchi:2004gf,Frezzotti:2005zm,Sint:2005qz}.
As was discussed in Ref.~\cite{Taniguchi:2004gf} the chirally twisted
boundary condition has advantages that the fermion determinant becomes
real and the mass term is introduced easier.
For domain-wall fermion the Pauli-Villars field can be treated in a
straightforward way by orbifolding.

\subsection{Orbifolding construction of chirally twisted boundary condition}

As will be discussed later the fermion determinant becomes real for even
numbers of flavours.
In this section we adopt two flavours case for instance.
We start from the massless orbifolded action \eqn{eqn:massless} and
introduce the twisted orbifolding by chirally rotating the fermion field
\begin{eqnarray}
\psi=e^{i\frac{\pi}{4}Q\tau^3}\psi',\quad
\bpsi=\bpsi'e^{-i\frac{\pi}{4}Q\tau^3},
\label{eqn:chiral-rot}
\end{eqnarray}
where $\tau^3$ is the Pauli matrix to act on flavour space and $Q$ is
the vector charge \eqn{eqn:Q} for chiral transformation.
In terms of the rotated field the orbifolded action is given by
\begin{eqnarray}
S_{\rm SF}=\sum\frac{1}{2}\bpsi'\wt{D}_{\rm dwf}^{\rm SF}\psi',\quad
\wt{D}_{\rm dwf}^{\rm SF}=\otPi_-D_{\rm dwf}\otPi_-,
\end{eqnarray}
where
\begin{eqnarray}
\otPi_\pm=\frac{1\pm\ovl{\Sigma}\tau^3}{2}
\end{eqnarray}
is a twisted orbifolding projection with time reflection operator
$\ovl{\Sigma}$ defined in \eqn{eqn:timeref}.

As was discussed in sub-section \ref{sec:orbifolding} the Dirac operator
has no index and the chiral transformation is not anomalous even for
Abelian case.
This formulation with twisted orbifolding projection is equivalent to
the original one for massless theory.
Here we notice that the twisted orbifolding operator
$\ovl{\Sigma}\tau^3$ commute with the massive domain-wall fermion Dirac
operator
\begin{eqnarray}
\left[\ovl{\Sigma}\tau^3,D_{\rm dwf}(m_f)\right]=0
\end{eqnarray}
since we adopted time reflection invariant gauge configuration.
We can extend this twisted formulation to massive theory
\begin{eqnarray}
S_{\rm dwf}^{\rm twist}=\sum\frac{1}{2}\bpsi\wt{D}_{\rm dwf}^{\rm SF}(m_f)\psi,
\quad
\wt{D}_{\rm dwf}^{\rm SF}=\otPi_-D_{\rm dwf}(m_f)\otPi_-.
\label{eqn:twisted-action}
\end{eqnarray}
%This corresponds to adopt chirally twisted mass term in the original
%orbifolded theory.
It is straightforward to introduce the Pauli-Villars field through
orbifolding
\begin{eqnarray}
S_{\rm PV}^{\rm twist}=\sum\frac{1}{2}\bphi\wt{D}_{\rm PV}^{\rm SF}\phi,\quad
\wt{D}_{\rm PV}^{\rm SF}=\otPi_-D_{\rm PV}\otPi_-
\label{eqn:PV}
\end{eqnarray}
since $D_{\rm PV}=D_{\rm dwf}(m_f=1)$ and is commutable with the
orbifolding operator.

The fermion fields satisfy the twisted orbifolding projection condition
\begin{eqnarray}
\otPi_+\psi=0,\quad
\bpsi\otPi_+=0,
\label{eqn:twisted-projection}
\end{eqnarray}
which brings the following boundary conditions
\begin{eqnarray}
&&
\otP_+\psi(\vec{x},0,s)=0,\quad
\otP_-\psi(\vec{x},N_T,s)=0
\\&&
\left(\bpsi\;\otP_+\right)(\vec{x},0,s)=0,\quad
\left(\bpsi\;\otP_-\right)(\vec{x},N_T,s)=0
\end{eqnarray}
with projection operator
\begin{eqnarray}
\otP_\pm=\frac{1\pm i\gamma_5\gamma_0P\tau^3}{2}.
\end{eqnarray}

In terms of the physical quark field the projection condition becomes
\begin{eqnarray}
&&
\left(P_L\delta_{s,1}+P_R\delta_{s,N_5}\right)
\left(\otPi_+\right)_{s,t}\psi(x,t)
%\nn\\&&
%=\frac{1}{2}\left(P_L\psi(x,1)+\Sigma\tau^3P_R\psi(x,N)\right)
%+\frac{1}{2}\left(P_R\psi(x,N)+\Sigma\tau^3P_L\psi(x,1)\right)
%\nn\\&&
%=\frac{1}{2}\left(1+\Sigma\tau^3\right)\left(P_L\psi(x,1)+P_R\psi(x,N)\right)
=\tPi_+q(x)=0,
\\&&
\bpsi(x,t)\left(\otPi_+\right)_{t,s}
\left(\delta_{s,N_5}P_L+\delta_{s,1}P_R\right)
%\nn\\&&
%=\frac{1}{2}\left(\bpsi(x,N)P_L+\bpsi(x,1)\Sigma\tau^3P_L\right)
%+\frac{1}{2}\left(\bpsi(x,1)P_R+\bpsi(x,N)\Sigma\tau^3P_R\right)
%\nn\\&&
%=\left(\bpsi(x,N)P_L+\bpsi(x,1)P_R\right)\frac{1}{2}\left(1+\Sigma\tau^3\right)
=\bq(x)\tPi_+=0,
\\&&
\tPi_\pm=\frac{1\pm\Sigma\tau^3}{2},\quad
\Sigma=i\gamma_5\gamma_0R,
\end{eqnarray}
where $\Sigma$ is the time reflection operator in four dimensions.
The boundary condition for the physical quark field is
\begin{eqnarray}
&&
\tP_+q(x)|_{x_0=0}=0,\quad
\tP_-q(x)|_{x_0=N_T}=0,
%\label{eqn:DBC1}
\\&&
\bq(x)\tP_+|_{x_0=0}=0,\quad
\bq(x)\tP_-|_{x_0=N_T}=0,
%\label{eqn:DBC2}
\\&&
\tP_\pm=\frac{1\pm i\gamma_5\gamma_0\tau^3}{2}.
\end{eqnarray}

We have two comments.
The orbifolded Dirac operator with twisted projection has a following
Hermiticity relation
\begin{eqnarray}
\wt{D}_{\rm SF}(m)^\dagger
=\gamma_5\tau^{1,2}\wt{D}_{\rm SF}(m)\gamma_5\tau^{1,2},
\end{eqnarray}
which is also the same for the orbifolded Pauli-Villars Dirac operator.
The $SU(2)$ flavour symmetry is broken to $U(1)_V\times U(1)_3$ as in
the chirally twisted mass QCD.

\subsection{Free propagator}

The original theory before orbifolding has a mass gap proportional to
$1/T$ because of anti-periodicity in temporal direction.
This property is robust against orbifolding process and survive in the
twisted orbifolding formulation.
We will check this property at tree level by using propagator.

The fermion propagator is defined as an inverse of the orbifolded Dirac
operator in a sub-space
\begin{eqnarray}
&&
\wt{G}^{\rm SF}_{\rm dwf}(x,y;s,t)
=2\left(\wt{D}^{\rm SF}_{\rm dwf}\right)^{-1}_{x,y;s,t}
=2\left(\otPi_-\frac{1}{D_{\rm dwf}}\otPi_-\right)_{x,y;s,t},
\\&&
\wt{D}^{\rm SF}_{\rm dwf}\left(\wt{D}^{\rm SF}_{\rm dwf}\right)^{-1}
=\left(\wt{D}^{\rm SF}_{\rm dwf}\right)^{-1}\wt{D}^{\rm SF}_{\rm dwf}
=\otPi_-.
\end{eqnarray}

At tree level this propagator can be written in a simple form as
\begin{eqnarray}
\wt{G}^{\rm SF}_{\rm dwf}(x,y;s,t)&=&
\frac{1}{N_L^3}\sum_{\vec{p}}e^{i\vec{p}\left(\vec{x}-\vec{y}\right)}
\wt{G}^{\rm SF}_{\rm dwf}(\vec{p};x_0,y_0;s,t),
\\
\wt{G}^{\rm SF}_{\rm dwf}(\vec{p};x_0,y_0;s,t)&=&
\frac{1}{2aN_T}\sum_{n=-N_T+1}^{N_T}
\left(\frac{1}{D_{\rm dwf}(p)}\right)_{s,t'}
\biggl\{
 \left(e^{ip_0(x_0-y_0)}-e^{ip_0(x_0+y_0)}\right)\left(\otP_+\right)_{t',t}
\nn\\&&\quad
+\left(e^{ip_0(x_0-y_0)}+e^{ip_0(x_0+y_0)}\right)\left(\otP_-\right)_{t',t}
\biggr\}.
\end{eqnarray}
$D_{\rm dwf}(p)$ is the domain-wall fermion Dirac operator in momentum
space without orbifolding projection, whose inverse is given in appendix
C.
We notice that the temporal momentum $p_0$ satisfies the quantization
condition \eqn{eqn:p0} and there is no extra fermion zero mode.

The physical quark propagator is given by selecting the contribution
from the boundary fields in fifth direction
\begin{eqnarray}
\wt{G}^{\rm SF}_{\rm quark}(x,y)&=&
\left(P_L\delta_{s,1}+P_R\delta_{s,N_5}\right)
\wt{G}^{\rm SF}_{\rm dwf}(x,y;s,t)
\left(\delta_{t,N_5}P_L+\delta_{t,1}P_R\right)
\nn\\&=&
2\left(\tPi_-G_{\rm quark}\tPi_-\right)_{x,y},
%\label{eqn:qprop}
\end{eqnarray}
where $G_{\rm quark}(x,y)$ is the physical quark propagator in
$2N_T\times N_L^3$ space-time without any projection.
Following Dirichlet boundary conditions
\begin{eqnarray}
&&
\tP_+G^{\rm SF}_{\rm quark}(x,y)|_{x_0=0}=0,\quad
\tP_-G^{\rm SF}_{\rm quark}(x,y)|_{x_0=N_T}=0,
\\&&
G^{\rm SF}_{\rm quark}(x,y)|_{y_0=0}\tP_+=0,\quad
G^{\rm SF}_{\rm quark}(x,y)|_{y_0=N_T}\tP_-=0
\end{eqnarray}
are satisfied for this quark propagator.
By ignoring sub-leading terms in $e^{-N_5}$ the propagator takes the
following form at tree level
\begin{eqnarray}
a^3\sum_{\vec{x}}e^{-i\vec{p}\left(\vec{x}-\vec{y}\right)}
G^{\rm SF}_{\rm quark}(x,y)&=&
\frac{1}{2aN_T}\sum_{n=-N_T+1}^{N_T}
\left(\frac{i\gamma_\mu \sin p_\mu}{1-e^{\alpha}W(p)}\right)
e^{ip_0x_0}
\nn\\&&\times
\left\{\left(e^{-ip_0y_0}-e^{ip_0y_0}\right)\tP_+
+\left(e^{-ip_0y_0}+e^{ip_0y_0}\right)\tP_-\right\}.
\end{eqnarray}
We emphasize that the physical quark has a gap proportional to $1/T$
because of the anti-periodicity \eqn{eqn:p0}.
This formulation satisfies one of the requirement.

\subsection{Surface term}

In this subsection we consider a twisted orbifolding symmetry and
introduce a coupling to the boundary source field (surface term) as a
symmetry breaking term.
The orbifolded action \eqn{eqn:twisted-action} is invariant under the
following twisted orbifolding transformation
\begin{eqnarray}
&&
\delta\left(\otPi_-\psi\right)(x,s)=\alpha\left(\otPi_-\psi\right)(x,s),
\quad
\delta\left(\bpsi\otPi_-\right)(x,s)=-\alpha\left(\bpsi\otPi_-\right)(x,s),
\label{eqn:inftr3}
\end{eqnarray}
where remaining degrees of freedom $\otPi_+\psi$ and $\bpsi\otPi_+$ are
kept intact.
The boundary source fields are elements of $\otPi_+\psi$ and
$\bpsi\otPi_+$.

We define a surface term as an orbifolding symmetry breaking term, which
is consistent with parity
\begin{eqnarray}
\psi(x,s)\to\gamma_0P_{st}\tau^{1,2}\psi(-\vec{x},x_0,t),\quad
\bpsi(x,s)\to\bpsi(-\vec{x},x_0,t)\gamma_0P_{ts}\tau^{1,2},
\end{eqnarray}
charge conjugation
\begin{eqnarray}
\psi(x,s)\to CP_{st}\tau^{1,2}\bpsi^T(x,t),\quad
\bpsi(x,s)\to \psi^T(x,t)\left(-C^{-1}\right)P_{ts}\tau^{1,2},\quad
C=\gamma_2\gamma_0
\end{eqnarray}
and
\begin{eqnarray}
\psi(x,s)\to C(PQ)_{st}\bpsi^T(x,t),\quad
\bpsi(x,s)\to \psi^T(x,t)\left(-C^{-1}\right)(PQ)_{ts},\quad
C=\gamma_2\gamma_0
\end{eqnarray}
and vector $U(1)_3$ symmetry
\begin{eqnarray}
\delta\psi(x,s)=\beta\tau^3\psi(x,s),\quad
\delta\bpsi(x,s)=-\beta\bpsi(x,s)\tau^3.
\end{eqnarray}
of the orbifolded domain-wall fermion.
Here we modified the parity and the charge conjugation transformation
to be consistent with the twisted orbifolding projection.

Using the orbifolding projection \eqn{eqn:twisted-projection}
the orbifolding transformation \eqn{eqn:inftr3} is shown to be a
``chiral'' transformation at the boundary in which a half of degrees is
rotated
\begin{eqnarray}
&&
\delta\left(\otP_-\psi\right)(\vec{x},0,s)
=\alpha\left(\otP_-\psi\right)(\vec{x},0,s),
\\&&
\delta\left(\bpsi\otP_-\right)(\vec{x},0,s)
=-\alpha\left(\bpsi\otP_-\right)(\vec{x},0,s),
\\&&
\delta\left(\otP_+\psi\right)(\vec{x},N_T,s)
=\alpha\left(\otP_+\psi\right)(\vec{x},N_T,s),
\\&&
\delta\left(\bpsi\otP_+\right)(\vec{x},N_T,s)
=-\alpha\left(\bpsi\otP_+\right)(\vec{x},N_T,s)
\end{eqnarray}
and is a vector $U(1)$ transformation in the bulk $0<x_0<N_T$
\begin{eqnarray}
\delta\psi(\vec{x},x_0,s)=\alpha\psi(\vec{x},x_0,s),\quad
\delta\bpsi(\vec{x},x_0,s)=-\alpha\bpsi(\vec{x},x_0,s).
\end{eqnarray}
The symmetry should be broken at the boundary.

We introduce boundary source fields as a component of projected out
degrees of freedom
\begin{eqnarray}
&&
\lambda(\vec{x},s)=\otP_+\psi(\vec{x},0,s),\quad
\lambda'(\vec{x},s)=\otP_-\psi(\vec{x},N_T,s),
\\&&
\blambda(\vec{x},s)=\left(\bpsi\;\otP_+\right)(\vec{x},0,s),\quad
\blambda'(\vec{x},s)=\left(\bpsi\;\otP_-\right)(\vec{x},N_T,s).
\end{eqnarray}
The orbifolding symmetry breaking term takes the form
\begin{eqnarray}
S_{\rm breaking}&=&
\blambda(\vec{x},s)\wt{O}_{st}\otP_-\psi(\vec{x},0,t)
+\left(\bpsi\;\otP_-\right)(\vec{x},0,s)\wt{O}_{st}\lambda(\vec{x},t)
\nn\\&&
+\blambda'(\vec{x},s)\wt{O}_{st}\otP_+\psi(\vec{x},N_T,t)
+\left(\bpsi\;\otP_+\right)(\vec{x},N_T,s)\wt{O}_{st}\lambda'(\vec{x},t),
\end{eqnarray}
where $\wt{O}$ is a local operator which anti-commute with
$i\gamma_5\gamma_0P\tau^3$.
Candidates of $\wt{O}$ which is consistent with the parity, charge
conjugation and $U(1)_3$ symmetries are
$PQ\tau^3$, $K(u)Q\tau^3$ and $\wt{K}(u)Q\tau^3$, where $K$ and $\wt{K}$
are defined in \eqn{eqn:K} \eqn{eqn:wtK}.

As in the previous section we adopt $K(1)Q$ for the surface term to
couple only to the physical quark field
\begin{eqnarray}
S_{\rm surface}&=&
-a^3\sum_{\vec{x}}\Bigl(
\blambda(\vec{x},s)(K(1)Q)_{st}\tau^3\otP_-\psi(\vec{x},0,t)
+\left(\bpsi\;\otP_-\right)(\vec{x},0,s)(K(1)Q)_{st}\tau^3\lambda(\vec{x},t)
\nn\\&&
+\blambda'(\vec{x},s)(K(1)Q)_{st}\tau^3\otP_+\psi(\vec{x},N_T,t)
+\left(\bpsi\;\otP_+\right)(\vec{x},N_T,s)(K(1)Q)_{st}\tau^3\lambda'(\vec{x},t)
\Bigr)
\nn\\&=&
a^3\sum_{\vec{x}}\Bigl(
-\left.\ovl{\rho}(\vec{x})\gamma_5\tau^3\wt{P}_-q(x)\right|_{x_0=0}
-\left.\bq(x)\wt{P}_-\gamma_5\tau^3\rho(\vec{x})\right|_{x_0=0}
\nn\\&&
-\left.\ovl{\rho}'(\vec{x})\gamma_5\tau^3\wt{P}_+q(x)\right|_{x_0=N_T}
-\left.\bq(x)\wt{P}_+\gamma_5\tau^3\rho'(\vec{x})\right|_{x_0=N_T}
\Bigr).
\label{eqn:surface2}
\end{eqnarray}
$\rho$ and $\brho$ are boundary source fields for the physical quark
fields
\begin{eqnarray}
&&
\wt{P}_+q(x)|_{x_0=0}=\rho(\vec{x}),\quad
\wt{P}_-q(x)|_{x_0=N_T}=\rho'(\vec{x}),
\\&&
\bq(x)\wt{P}_+|_{x_0=0}=\brho(\vec{x}),\quad
\bq(x)\wt{P}_-|_{x_0=N_T}=\brho'(\vec{x}).
\end{eqnarray}

Although this surface term is not a general symmetry breaking term,
we also expect that quantum corrections can be renormalized into a shift
of physical operators and physical quark source fields $\rho$, $\brho$,
$\rho'$ and $\brho'$ if we consider Green functions constructed with
physical quark operators only.

\subsection{Effective action of the domain-wall fermion}

For the twisted orbifolding formulation of finite volume field theory
the Pauli-Villars field is introduced directly as in \eqn{eqn:PV}.
Total contributions from fermion and Pauli-Villars field is
\begin{eqnarray}
\det_{\ovl{\wt{\cal H}}_+}\otPi_-\frac{1}{D_{\rm PV}}D_{\rm dwf}(m_f)\otPi_-,
\end{eqnarray}
where the determinant is defined in a sub-space
$\ovl{\wt{\cal H}}_+=\left\{\psi|\otPi_+\psi=0\right\}$ of
eigenfunctions.
In this sub-section we will show that this determinant is equivalent to
that of the overlap Dirac operator with twisted orbifolding
\cite{Taniguchi:2004gf}
\begin{eqnarray}
\det_{\ovl{\wt{\cal H}}_+}\otPi_-\frac{1}{D_{\rm PV}}D_{\rm dwf}(m_f)\otPi_-
=\det_{{\wt{\cal H}}_+}\wt{\Pi}_-D_{\rm OD}(m_f)\wt{\Pi}_-.
\end{eqnarray}

For this purpose we adopt the Schur decomposition procedure for the
effective Dirac operator \cite{Edwards:2000qv,Edwards:2005an}.
Statement of the Schur decomposition is that the overlap Dirac operator
is given as a Schur complement of the domain-wall fermion Dirac operator
divided by the Pauli-Villars Dirac operator
\begin{eqnarray}
\frac{1}{D_{\rm PV}}D_{\rm dwf}(m_f)
={\cal P}U^{-1}(1)D_{\rm OD}^{(5)}(m_f)U(m_f){\cal P}^\dagger.
\end{eqnarray}
Here ${\cal P}$, $U(m_f)$ and $D_{\rm OD}^{(5)}(m_f)$ are matrices in
fifth dimension and their explicit forms for $N_5=6$ case are given by
\begin{eqnarray}
&&
{\cal P}=\pmatrix{
P_R&   &   &   &   &P_L\cr
P_L&P_R&   &   \cr
   &P_L&P_R&   \cr
   &   &P_L&P_R\cr
   &   &   &P_L&P_R\cr
   &   &   &   &P_L&P_R\cr}
=P_R+\Omega^-(-1)P_L,
\\&&
U(m_f)=\pmatrix{
1& & & & &-T\left(P_L-m_fP_R\right)\cr
 &1& & & &-T^2\left(P_L-m_fP_R\right)\cr
 & &1& & &-T^3\left(P_L-m_fP_R\right)\cr
 & & &1& &-T^4\left(P_L-m_fP_R\right)\cr
 & & & &1&-T^5\left(P_L-m_fP_R\right)\cr
 & & & & &1\cr},
\\&&
D_{\rm OD}^{(5)}(m_f)
=\pmatrix{1\cr&1\cr&&1\cr&&&1\cr&&&&1\cr&&&&&D_{\rm OD}(m_f)\cr},
\end{eqnarray}
where $D_{\rm OD}(m_f)$ is a truncated four dimensional massive overlap
Dirac operator
\begin{eqnarray}
D_{\rm OD}(m_f)=
\frac{1}{2}\left(1+\gamma_5S\right)
+m_f\left(1-\frac{1}{2}\left(1+\gamma_5S\right)\right)
\end{eqnarray}
with the same definition for $S$ in \eqn{eqn:truncated}.
The truncated Dirac operator turns out to be the ordinary overlap Dirac
operator \eqn{eqn:massiveOD} in $N_5\to\infty$ limit.
$\Omega^-(m_f)$ is a hopping operator in fifth direction
\eqn{eqn:omega}.
So we have
\begin{eqnarray}
\det\frac{1}{D_{\rm PV}}D_{\rm dwf}(m_f)=\det D_{\rm OD}(m_f)
\end{eqnarray}
for ordinary domain-wall fermion system.

We start from the orbifolded domain-wall fermion Dirac operator divided
by the Pauli-Villars Dirac operator
\begin{eqnarray}
D_{\rm SF}^{(5)}=\otPi_-\frac{1}{D_{\rm PV}}D_{\rm dwf}(m_f)\otPi_-
=\otPi_-{\cal P}U^{-1}(1)D_{\rm OD}^{(5)}(m_f)U(m_f){\cal P}^\dagger\otPi_-.
\end{eqnarray}
We consider multiplication of the projection operator on unitary matrix
${\cal P}$ and we have
\begin{eqnarray}
\otPi_-{\cal P}=\otPi_-{\cal P}\wh{\oP}_-,\quad
\wh{\oP}_\pm=\frac{1\pm P\Omega^-(-1)\Sigma\tau^3}{2}.
\end{eqnarray}
We notice that a matrix in the projection $\wh{\oP}_\pm$ has a following
form
\begin{eqnarray}
P\Omega^-(-1)&=&\Omega^+(-1)P
=\pmatrix{P_{(N_5-1)}&0\cr 0&1\cr},
\end{eqnarray}
where $P_{(N_5-1)}$ is a $(N_5-1)\times(N_5-1)$ matrix of the form
\begin{eqnarray}
P_{(N_5-1)}=\pmatrix{
0 & 0 & 0 & 0 & 1 \cr
0 & 0 & 0 & 1 & 0 \cr
0 & 0 & 1 & 0 & 0 \cr
0 & 1 & 0 & 0 & 0 \cr
1 & 0 & 0 & 0 & 0 \cr},\quad
(N_5=6).
\end{eqnarray}
The projection operator $\wh{\oP}_\pm$ is written as a direct sum of two
projections
\begin{eqnarray}
\wh{\oP}_\pm=\pmatrix{\otPi_\pm^{(N_5-1)}\cr & \wt{\Pi}_\pm}
\end{eqnarray}
where
\begin{eqnarray}
\otPi_\pm^{(N_5-1)}=\frac{1\pm P_{(N_5-1)}\Sigma\tau^3}{2}
\end{eqnarray}
is a projection operator in $N_5-1$ sub-space.

Taking into account the explicit form of the matrix $U(m_f)$ its
determinant multiplied by the projection becomes
\begin{eqnarray}
&&
\det_{(+{\rm subspace})}U(m)\wh{\oP}_-
=\det_{(+{\rm subspace})}\wh{\oP}_-U(m)\wh{\oP}_-
=\det_{(+{\rm subspace})}\pmatrix{\otPi_-^{(N_5-1)}\cr & \wt{\Pi}_-}
=1,
\\&&
\det_{(+{\rm subspace})}\wh{\oP}_-U^{-1}(m)
=\det_{(+{\rm subspace})}\wh{\oP}_-U^{-1}(m)\wh{\oP}_-
=\det_{(+{\rm subspace})}\pmatrix{\otPi_-^{(N_5-1)}\cr & \wt{\Pi}_-}
=1.
\nn\\
\end{eqnarray}
Substituting this relation determinant of the total Dirac operator is
equivalent to that of the orbifolded overlap Dirac operator
\begin{eqnarray}
\det_{\ovl{\wt{\cal H}}_+}D_{\rm SF}^{(5)}&=&
\det_{\ovl{\wt{\cal H}}_+}\wt{\oP}_-{\cal P}\wh{\oP}_-U^{-1}(1)
\wh{\oP}_-D_{\rm OD}^{(5)}(m_f)\wh{\oP}_-U(m_f)\wh{\oP}_-
{\cal P}^\dagger\wt{\oP}_-
\nn\\&=&
\det_{(+{\rm subspace})}\left(\wh{\oP}_-D_{\rm OD}^{(5)}(m_f)\wh{\oP}_-\right)
\nn\\&=&
\det_{(+{\rm subspace})}
\pmatrix{\wt{\oP}_-^{(N_5-1)}\cr & \wt{\Pi}_-}
\pmatrix{1_{(N_5-1)}\cr &D_{\rm OD}(m_f)}
\pmatrix{\wt{\oP}_-^{(N_5-1)}\cr & \wt{\Pi}_-}
\nn\\&=&
\det_{\wt{\cal H}_+}\wt{\Pi}_-D_{\rm OD}(m_f)\wt{\Pi}_-
\end{eqnarray}
and we get expected result.

At last we have a comment on Hermiticity.
The five dimensional total Dirac operator $D_{\rm SF}^{(5)}$ has a
following Hermiticity relation
\begin{eqnarray}
{D_{\rm SF}^{(5)}}^\dagger
=\gamma_5\tau^{1,2}D_{\rm SF}^{(5)}\gamma_5\tau^{1,2}
\end{eqnarray}
and its determinant is real.
Since our domain-wall fermion Dirac operator does not have index
the chiral rotation \eqn{eqn:chiral-rot} is well defined even for single
flavour case and we can define a single flavour orbifolded Dirac
operator as
\begin{eqnarray}
D_{\rm SF}^{\rm single}=
\frac{1-\ovl{\Sigma}}{2}\frac{1}{D_{\rm PV}}D_{\rm dwf}(m_f)
\frac{1-\ovl{\Sigma}}{2}.
\end{eqnarray}
However we do not have a Hermiticity relation for this Dirac operator
and the determinant is not shown to be real.
We may need even numbers of flavours to avoid this problem.

\reseteqnum
%%%%%%%%%%%%%%%%%%%%%%%%%%%%%Section 4%%%%%%%%%%%%%%%%%%%%%%%%%%%%%%%
\section{Conclusion}
\label{sec:conclusion}

In this paper the orbifolding formulation of the finite volume field
theory is applied to the domain-wall fermion.
In order to reproduce the proper SF Dirichlet boundary condition we need
both the time reflection and the chiral symmetries.
Application of this procedure to fermionic part is straightforward
because of good chiral symmetry of the domain-wall fermion.
Since there is no chiral symmetry for the Pauli-Villars field it is
introduced by using the effective Dirac operator to reproduce the SF
overlap Dirac operator.
The surface term is given as an external source field to break the
orbifolding symmetry.

The SF Dirichlet boundary condition may not be the unique choice to
define a finite volume field theory suitable for renormalization scheme.
A finite volume field theory with chirally twisted boundary condition is
also proposed.
Time reflection symmetry is enough to reproduce the twisted boundary
condition by orbifolding.
We can treat the fermionic part and the Pauli-Villars field in an equal
footing.
We have a $\gamma_5$ Hermiticity relation for the orbifolded Dirac
operator and the total determinant is real.
This formulation is applicable to two flavours dynamical simulation.

\appendix
\reseteqnum
%%%%%%%%%%%%%%%%%%%%%%%%%%%%%Section 4%%%%%%%%%%%%%%%%%%%%%%%%%%%%%%%
\section{Effective action of chiral symmetric Dirac operator}

In this appendix we derive the effective Dirac operator of the physical
quark field for an action with the chiral symmetric Dirac operator
\eqn{eqn:symmetric}.
Four dimensional part of the symmetric Dirac operator is the same as
the ordinary Dirac operator \eqn{eqn:dirac-op}.
Hopping term of this Dirac operator into the fifth direction takes the
form
\begin{eqnarray}
P_L\Omega^+(m_f=0)+P_R\Omega^-(m_f=0)
=\pmatrix{
0  &P_L& 0 & 0 & 0 & 0 \cr
P_R& 0 &P_L& 0 & 0 & 0 \cr
0  &P_R& 0 & 0 & 0 & 0 \cr
0  & 0 & 0 & 0 &P_L& 0 \cr
0  & 0 & 0 &P_R& 0 &P_L\cr
0  & 0 & 0 & 0 &P_R& 0 \cr}
\end{eqnarray}
for massless case.
If there were no quark mass this Dirac operator is equivalent to
two independent domain-wall fermion with half fifth dimensional size
of $N_5/2$.
It is easily shown that there are two extra zero mode (doublers) at the
middle boundary $s=\frac{N_5}{2}$ and $s=\frac{N_5}{2}+1$ related to the
exact chiral symmetry at finite $N_5$.

The physical quark fields may be defined in the same manner as
\eqn{eqn:q} and \eqn{eqn:bq}.
% and the Pauli-Villars field can be introduced in the ordinary way.
We can integrate out the bulk field other than $q$ and $\bq$ according
to Ref.~\cite{Kikukawa:1999sy,Neuberger:1997bg,Edwards:2000qv}.
We start by writing the fermion field as a vector in fifth direction and
chirality.
For $N_5=6$ we have
\begin{eqnarray}
&&
\Psi^T=\pmatrix{
\psi_{1R}&\psi_{1L}&\psi_{2R}&\psi_{2L}&
\psi_{3R}&\psi_{3L}&\psi_{4R}&\psi_{4L}&
\psi_{5R}&\psi_{5L}&\psi_{6R}&\psi_{6L}\cr},
\nn\\&&
\bPsi=\pmatrix{
\bpsi_{1L}&\bpsi_{1R}&\bpsi_{2L}&\bpsi_{2R}&
\bpsi_{3L}&\bpsi_{3R}&\bpsi_{4L}&\bpsi_{4R}&
\bpsi_{5L}&\bpsi_{5R}&\bpsi_{6L}&\bpsi_{6R}\cr},
\nn
\end{eqnarray}
where
\begin{eqnarray}
\psi_{R/L}=P_{R/L}\psi,\quad
\bpsi_{R/L}=\bpsi P_{L/R}.
\end{eqnarray}
Then we change variable as
\begin{eqnarray}
&&
\Psi^{'T}=\pmatrix{
\psi_{1L}&\psi_{1R}&\psi_{2L}&\psi_{2R}&
\psi_{3L}&\psi_{3R}&\psi_{4L}&\psi_{4R}&
\psi_{5L}&\psi_{5R}&\psi_{6L}&\psi_{6R}\cr},
\nn\\&&
\bPsi'=\pmatrix{
\bpsi_{1R}&\bpsi_{2L}&\bpsi_{2R}&\bpsi_{3L}&\bpsi_{3R}&\bpsi_{1L}&
\bpsi_{4R}&\bpsi_{5L}&\bpsi_{5R}&\bpsi_{6L}&\bpsi_{6R}&\bpsi_{4L}\cr}.
\nn
\end{eqnarray}
The Dirac operator is written as follows in terms of the primed field
\begin{eqnarray}
D_{\rm dwf}=\pmatrix{
\alpha&\beta  \cr
      &\alpha&\beta  \cr
\beta_0&&\alpha_0&&&\cr
      &&&\alpha&\beta  \cr
      &&&&\alpha&\beta  \cr
&&&\beta_0&&\alpha_0\cr},
\end{eqnarray}
where
\begin{eqnarray}
&&
\alpha=\pmatrix{ B &-C^\dagger\cr   &-1 \cr},\quad
\alpha_0=P_R\alpha,
\\&&
\beta=\pmatrix{-1 &   \cr C & B \cr},\quad
\beta_0=P_L\beta,
\\&&
C_{xy}=\sigma_\mu\frac{1}{2}
\left(\delta_{x+\mu,y}U_\mu(x)-\delta_{x-\mu,y}U_\mu^\dagger(y)\right),
\\&&
B_{xy}=(1-M)\delta_{xy}
-\frac{1}{2}\left(\delta_{x+\mu,y}U_\mu(x)+\delta_{x-\mu,y}U_\mu^\dagger(y)
-2\delta_{xy}\right),
\\&&
\gamma_\mu=\pmatrix{&\sigma_\mu\cr\sigma^\dagger_\mu\cr}.
\end{eqnarray}

We integrate out all the fields except for the physical quark field
\begin{eqnarray}
&&
q(x)=P_L\psi(x,1) +P_R\psi(x,N_5)
=P_L\gamma_0\psi'(x,1)+P_R\gamma_0\psi'(x,N_5),
\\&&
\bq(x)=\bpsi(x,1)P_R +\bpsi(x,N_5)P_L
=\bpsi'(x,\frac{N_5}{2})\gamma_0P_R+\bpsi'(x,N_5)\gamma_0P_L
\end{eqnarray}
according to Ref.~\cite{Kikukawa:1999sy}.
Result is given as a full quark propagator
\begin{eqnarray}
\vev{q\bq}=
\frac{1}{2}\left(\frac{1}{D_{\rm eff}}
-\gamma_5\frac{1}{D_{\rm eff}}\gamma_5\right)
=\frac{1}{D_{\rm eff}^{\rm sym}}.
\end{eqnarray}
Here $D_{\rm eff}$ is the truncated effective Dirac operator
\eqn{eqn:truncated} with half size of fifth dimensional length
\begin{eqnarray}
D_{\rm eff}=\frac{1+\gamma_5S}{1-\gamma_5S},\quad
S=\frac{1-T^{\frac{N_5}{2}}}{1+T^{\frac{N_5}{2}}}.
\end{eqnarray}
Transfer matrix is given by
\begin{eqnarray}
T=\gamma_5\gamma_0\left(-\alpha\beta^{-1}\right)\gamma_0\gamma_5
=\frac{1-H'}{1+H'}.
\end{eqnarray}
The full quark propagator anti-commutes with $\gamma_5$ even at finite
$N_5$.
In $N_5\to\infty$ limit the effective Dirac operator $D_{\rm eff}$
anti-commutes with $\gamma_5$ exactly and the effective Dirac operator
$D_{\rm eff}^{\rm sym}$ with symmetric construction becomes the same as
that of the ordinary domain-wall fermion $D_{\rm eff}$.

We introduce the Pauli-Villars field in the same manner with the Dirac
operator
\begin{eqnarray}
D_{\rm PV}^{\rm sym}=D_{\rm dwf}(m_f=1)+X.
\end{eqnarray}
The effective action of the physical quark field $q$, $\bq$ and the
physical Pauli-Villars field $Q$, $\bQ$ is given by
\begin{eqnarray}
S_{\rm eff}=\bq D_{\rm eff}^{\rm sym}q
+\bQ\left(D_{\rm eff}^{\rm sym}+1\right)Q.
\end{eqnarray}
The overlap Dirac operator is given to reproduce the same determinant as
the effective action
\begin{eqnarray}
D_{\rm OD}^{\rm sym}=\frac{D_{\rm eff}^{\rm sym}}{D_{\rm eff}^{\rm sym}+1}.
\end{eqnarray}
Because of exact chiral symmetry of $D_{\rm eff}^{\rm sym}$ the overlap
Dirac operator $D_{\rm OD}^{\rm sym}$ satisfies the Ginsparg-Wilson
relation even at finite $N_5$.

Compensation of the exact chirality at finite $N_5$ is a non-locality in
the overlap Dirac operator, which comes from the extra zero mode in the
middle of fifth direction.
However we can show that the non-locality is exponentially small in
$N_5$ and disappears in $N_5\to\infty$ limit.
In order to extract the non-locality we define explicit breaking term of
the chiral symmetry of the ordinary effective Dirac operator
\eqn{eqn:truncated} or the truncated overlap Dirac operator at finite
$N_5$
\begin{eqnarray}
\delta_{N_5}=\gamma_5\frac{1}{D_{\rm eff}(N_5)}
+\frac{1}{D_{\rm eff}(N_5)}\gamma_5
=\gamma_5\frac{1}{D_{\rm OD}(N_5)}+\frac{1}{D_{\rm OD}(N_5)}\gamma_5-2\gamma_5.
\end{eqnarray}
The chiral symmetric effective Dirac operator is re-written as
\begin{eqnarray}
\frac{1}{D_{\rm eff}^{\rm sym}}
=\frac{1}{D_{\rm eff}}-\frac{1}{2}\gamma_5\delta_{\frac{N_5}{2}},
\end{eqnarray}
where we used a fact that the breaking term commutes with $\gamma_5$
\begin{eqnarray}
\left[\delta_{N_5},\gamma_5\right]=0.
\end{eqnarray}
The chiral symmetric overlap Dirac operator is given in a following form
\begin{eqnarray}
D_{\rm OD}^{\rm sym}
=\frac{1}{1-\frac{1}{2}D_{\rm OD}\gamma_5\delta_{\frac{N_5}{2}}}D_{\rm OD}.
\end{eqnarray}
$D_{\rm OD}$ in denominator may bring a non-local factor into the
overlap Dirac operator.
However as was shown in Ref.~\cite{Kikukawa:1999sy}
$\delta_{N_5}$ is exponentially small in $N_5$.
The physical part of the chiral symmetric Dirac operator
$D_{\rm dwf}^{\rm sym}$ coincides with that of the ordinary Dirac
operator in $N_5\to\infty$ limit.

\reseteqnum
\section{Folding of temporal direction}

In our formulation with the orbifolding \eqn{eqn:orbifolding} fermion
fields in negative time $-N_T<x_0<0$ can be written in term of
those in the positive region
\begin{eqnarray}
\psi(\vec{x},-x_0,s)=\left(\ovl{\Gamma}\right)_{s,t}\psi(\vec{x},x_0,t),\quad
\ovl{\Gamma}=\gamma_0\gamma_5PQ.
\label{eqn:identify}
\end{eqnarray}
%For the boundary fields the orbifolding condition \eqn{eqn:orbifolding}
%is equivalent to Dirichlet boundary conditions
%\begin{eqnarray}
%&&
%\left(\ovl{P}_-\right)_{s,t}\psi(\vec{x},0,t)=0,\quad
%\left(\ovl{P}_+\right)_{s,t}\psi(\vec{x},N_T,t)=0,
%\\&&
%\bpsi(\vec{x},0,t)\left(\ovl{P}_-\right)_{t,s}=0,\quad
%\psi(\vec{x},N_T,t)\left(\ovl{P}_+\right)_{t,s}=0,
%\end{eqnarray}
%where $\ovl{P}_\pm$ is defined in \eqn{eqn:projection}.
Half of the field degrees of freedom can be eliminated explicitly by
folding the temporal axis into the non-negative range
$0\le x_0\le N_T$ together with the boundary condition
\eqn{eqn:projection-boundary1} \eqn{eqn:projection-boundary2}.

For this purpose we introduce four projection operators in temporal
direction
\begin{center}
\begin{tabular}{lllll}
$T_-$& & for& & $-N_T+1\le x_0\le-1$,\\
$T_0$& & for& & $x_0=0$,\\
$T_+$& & for& & $1\le x_0\le N_T-1$,\\
$T_T$& & for& & $x_0=N_T$,\\
\end{tabular}
\end{center}
which pick up the fermion fields on the corresponding region.
For instance 
\begin{eqnarray}
\left(T_+\right)_{x_0,y_0}\psi(y_0)&=&\left\{
\begin{array}{lll}
\psi(x_0) &{\rm for}& 1\le x_0\le N_T-1\\
0 &&{\rm otherwise}\\
\end{array}
\right..
\end{eqnarray}
Summing up four projection operators we have a unity
\begin{eqnarray}
1=T_-+T_0+T_++T_T
\end{eqnarray}
and $T_\alpha$'s have a projection property
\begin{eqnarray}
T_\alpha T_\beta=T_\alpha\delta_{\alpha,\beta}.
\end{eqnarray}
These projection operators satisfy the following relation with the time
reflection operator $R$
\begin{eqnarray}
RT_+=T_-R,\quad
RT_-=T_+R,\quad
RT_0=T_0R=T_0,\quad
RT_T=T_TR=-T_T.
\label{eqn:TR}
\end{eqnarray}

By using these properties we have an identity relation
\begin{eqnarray}
\oP_+&=&
\oP_+\left(T_++T_-+T_0+T_T\right)
\nn\\&=&
\oP_+\left(T_++T_0+T_T\right)\left(2T_++T_0+T_T\right)\oP_+
\end{eqnarray}
and the orbifolded action \eqn{eqn:massless} can be re-written in terms
of the fermion fields depending on the non-negative region only
\begin{eqnarray}
S=\sum_{\vec{x},\vec{y}}\sum_{x_0,y_0=0}^{N_T}\sum_{s,t}
{\bpsi}''(x,s)D^{\rm folded}_{\rm dwf}(x,y;s,t){\psi}''(y,t),
\end{eqnarray}
where ${\psi}''$ and ${\bpsi}''$ are defined as
\begin{eqnarray}
&&
{\psi}''(\vec{x},x_0,s)=
\left(\left(T_++T_0+T_T\right)\oP_+\right)_{x_0,y_0;s,t}\psi(\vec{x},y_0,t),
\\&&
{\bpsi}''(\vec{x},x_0,s)=
\bpsi(\vec{x},y_0,t)\left(\oP_+\left(T_++T_0+T_T\right)\right)_{y_0,x_0;t,s},
\end{eqnarray}
which have no dependence on negative time.
These fields can further be written as
\begin{eqnarray}
{\psi}''(\vec{x},x_0,s)&=&
\left(T_++T_0\ovl{P}_++T_T\ovl{P}_-\right)_{x_0,y_0;s,t}\psi(\vec{x},y_0,t),
\\
{\bpsi}''(\vec{x},x_0,s)&=&
\bpsi(\vec{x},y_0,t)\left(T_++\ovl{P}_+T_0+\ovl{P}_-T_T\right)_{y_0,x_0;t,s}
\end{eqnarray}
by using \eqn{eqn:TR} and identification \eqn{eqn:identify}.
There is no constraint on positive bulk fields.

The folded Dirac operator $\wt{D}^{\rm SF}_{\rm dwf}$ is given formally
as
\begin{eqnarray}
{D}^{\rm folded}_{\rm dwf}=
\frac{1}{2}\left(2T_++T_0+T_T\right)\oP_+D_{\rm dwf}\oP_+
\left(2T_++T_0+T_T\right).
\end{eqnarray}
This Dirac operator can be written in more explicit form by
using the orbifolding symmetry \eqn{eqn:symmetry} and the ultra local
property of the domain-wall fermion Dirac operator, with which we
eliminate the term like $T_+D_{\rm dwf}AT_+=T_+D_{\rm dwf}T_-A$
\begin{eqnarray}
{D}^{\rm folded}_{\rm dwf}&=&
\frac{1}{2}T_0\ovl{P}_+D_{\rm dwf}\ovl{P}_+T_0
+T_0\ovl{P}_+D_{\rm dwf}T_+
+T_+D_{\rm dwf}\ovl{P}_+T_0
+T_+D_{\rm dwf}T_+
\nn\\&&
+T_T\ovl{P}_-D_{\rm dwf}T_+
+T_+D_{\rm dwf}\ovl{P}_-T_T
+\frac{1}{2}T_T\ovl{P}_-D_{\rm dwf}\ovl{P}_-T_T
\\&=&
\pmatrix{
\ovl{P}_+\frac{D^{(3+1)}}{2}\ovl{P}_+&-\ovl{P}_+P_-U_0(0)\cr
-P_+\ovl{P}_+U_0^\dagger(0)&D^{(3+1)}&-P_-U_0(1)\cr
&-P_+U_0^\dagger(1)&D^{(3+1)}&-P_-U_0(2)\cr
&&-P_+U_0^\dagger(2)&D^{(3+1)}&-P_-\ovl{P}_-U_0(3)\cr
&&&-\ovl{P}_-P_+U_0^\dagger(3)&\ovl{P}_-\frac{D^{(3+1)}}{2}\ovl{P}_-}
\nn\\
\end{eqnarray}
where the matrix represents the Dirac operator in temporal direction
for $N_T=4$.
$D^{(3+1)}$ is the Dirac operator in spatial direction and the fifth
direction 
\begin{eqnarray}
D^{(3+1)}(x,y;s,t)&=&
\left(
 \frac{-1+\gamma_i}{2}U_i(x)\delta_{y_i,x_i+1}
+\frac{-1-\gamma_i}{2}U_i^\dagger(y)\delta_{y_i,x_i-1}\right)
\delta_{x_0,y_0}\delta_{s,t}
\nn\\&+&
\left(
\frac{-1+\gamma_5}{2}\Omega^+_{s,t}+\frac{-1-\gamma_5}{2}\Omega^-_{s,t}\right)
\delta_{x,y}
+(5-M)\delta_{x,y}\delta_{s,t}.
\end{eqnarray}
There is no constraint for the bulk region $1<x_0,y_0<N_T-1$,
which is nothing but ordinary domain-wall fermion Dirac operator.

We notice that the projection operator $\ovl{P}_\pm$ at the boundary
does not commute with the $\gamma_0$ chiral projection $P_\pm$.
If we consider an eigenvalue equation of this Dirac operator a zero mode
dumping solution
\begin{eqnarray}
\psi=P_-(1-M)^{x_0}+P_+(1-M)^{(N_T-x_0)}
\end{eqnarray}
in temporal direction, which have broken the chiral symmetry
``dynamically'' in a naive formulation, is forbidden by this boundary
term.

The fermion propagator is given as an inverse of the folded Dirac
operator 
\begin{eqnarray}
{G}^{\rm folded}_{\rm dwf}=
2\left(T_++T_0+T_T\right)\oP_+D_{\rm dwf}^{-1}\oP_+\left(T_++T_0+T_T\right),
\end{eqnarray}
where the inverse is defined in the ordinary meaning for the positive
bulk region $0<x_0<N_T$ and in terms of the projected sub-space at the
boundary 
\begin{eqnarray}
{D}^{\rm folded}_{\rm dwf}{G}^{\rm folded}_{\rm dwf}=
T_++\ovl{P}_+T_0+\ovl{P}_-T_T.
\end{eqnarray}

\reseteqnum
\section{Free fermion propagator}

Inverse of the massless domain-wall fermion Dirac operator in momentum
space is derived according to the procedure of
Ref.~\cite{Shamir:1993zy}.
In this appendix we omit derivation and give the result:
\begin{eqnarray}
\frac{1}{D_{\rm dwf}(p)}=
 \left(-i\gamma_\mu\sin p_\mu+W-\Omega^-\right)G_RP_L
+\left(-i\gamma_\mu\sin p_\mu+W-\Omega^+\right)G_LP_R,
\end{eqnarray}
where $\Omega$ and $W$ are defined in \eqn{eqn:omega} and \eqn{eqn:W}.
$G_R$ and $G_L$ are defined as
\begin{eqnarray}
&&
G_{\rm R} (s, t) = G^0(s-t) 
+A_{++} e^{\alpha (s+t)}
+A_{+-} e^{\alpha (s-t)}
+A_{-+} e^{\alpha (-s+t)}
+A_{--} e^{\alpha (-s-t)},
\\&&
G_{\rm L} (s, t) = G^0(s-t) 
+B_{++} e^{\alpha (s+t)}
+B_{+-} e^{\alpha (s-t)}
+B_{-+} e^{\alpha (-s+t)}
+B_{--} e^{\alpha (-s-t)},
\\&&
G^0 (s-t)=
C\left( e^{\alpha (N_5-|s-t|)} + e^{- \alpha (N_5-|s-t|)} \right)
\end{eqnarray}
with exponent and coefficients given by
\begin{eqnarray}
&&
\cosh\alpha = \frac{1+W^2+\sin^2p_\mu}{2|W|},
\\&&
C=\frac{1}{4W\sinh\alpha \sinh(\alpha N_5)},
\\&&
A_{++}=F(1-We^{-\alpha})(e^{-2\alpha N_5}-1),\quad
A_{--}=F(1-We^{\alpha})(1-e^{2\alpha N_5}),
\\&&
B_{++}=e^{-2\alpha(N_5+1)}A_{--},\quad
B_{--}=e^{2\alpha(N_5+1)}A_{++},
\\&&
A_{-+}=A_{+-}=B_{-+}=B_{+-}=FW(e^{\alpha}-e^{-\alpha}),
\\&&
F=\frac{C}{e^{\alpha N_5}(1-We^\alpha)-e^{-\alpha N_5}(1-We^{-\alpha})}.
\end{eqnarray}
This notation is valid for positive $W$ and for negative case we define
\begin{eqnarray}
e^{\pm\alpha}=\cosh\alpha\pm\sqrt{\cosh^2\alpha-1}
\end{eqnarray}
and flip their sign $e^{\pm\alpha}\to -e^{\pm\alpha}$ according to
${\rm sgn}(W)$.

The physical quark propagator in momentum space is defined by picking up
the boundary components 
\begin{eqnarray}
\vev{q(p)\bq(-p)}&=&
\left(P_L\delta_{s,1}+P_R\delta_{s,N_5}\right)
\left(\frac{1}{D_{\rm dwf}(p)}\right)_{s,t}
\left(\delta_{t,N_5}P_L+\delta_{t,1}P_R\right)
\nn\\&=&
-i\gamma_\mu\sin p_\mu G_R(N_5,N_5)+WG_R(1,N_5).
\end{eqnarray}
Ignoring the next to leading term in $N_5$ the quark propagator has a
simple form
\begin{eqnarray}
\vev{q(p)\bq(-p)}=\frac{i\gamma_\mu\sin p_\mu}{1-We^{\alpha}}.
\end{eqnarray}

\acknowledgments
I would like to thank M.~L\"uscher, S.~Aoki, O.~B\"ar, T.~Izubuchi,
Y.~Kikukawa and Y.~Kuramashi  for his valuable suggestions and
discussions.

%%%%%%%%%%%%%%%%%%%%%%%%%%%%%Section 1%%%%%%%%%%%%%%%%%%%%%%%%%%%%%%%

\end{document}